\documentclass[prl,aps, superscriptaddress,twocolumn, footinbib]{revtex4-2}
\usepackage{amsmath,amssymb}
\usepackage{graphicx}
\usepackage{wasysym}
\usepackage{amsfonts}
\usepackage{bm}
\usepackage{enumerate}
\usepackage{enumitem}
\usepackage{color}
\usepackage{xcolor}
\usepackage[resetlabels]{multibib}
\usepackage{epstopdf}
\usepackage{latexsym}
\usepackage[breaklinks,colorlinks = true,linkcolor = red,urlcolor=cyan,citecolor=red]{hyperref}
\usepackage[caption=false,singlelinecheck=false]{subfig}
\usepackage{hhline}
\usepackage{mathtools}
\usepackage{tabularray}
\usepackage{cancel}
\usepackage{tikz}

\usepackage{times}
\newcommand{\bea}{\begin{eqnarray}}
\newcommand{\eea}{\end{eqnarray}}
\newcommand{\be}{\begin{eqnarray}}
\newcommand{\ee}{\end{eqnarray}}
\newcommand{\bw}{\begin{widetext}}
\newcommand{\ew}{\end{widetext}}

\newcommand{\tcr}[1]{\textcolor{red}{#1}}

\newcommand\numberthis{\addtocounter{equation}{1}\tag{\theequation}}

\def\xo{\overline{X}}
\def\yo{\overline{Y}}
\def\zo{\overline{Z}}
\def\xu{\underline{X}}
\def\yu{\underline{Y}}
\def\zu{\underline{Z}}
\DeclareMathOperator{\len}{len}

\DeclareMathOperator{\coeff}{co}

\def\rd{0.425em}

\newcommand{\rcv}{
\begin{tikzpicture}[baseline=-0.5*\rd]
\draw[line width=0.5mm, red] (0,0) circle (1*\rd);
\end{tikzpicture}
}

\newcommand{\bsv}{
\begin{tikzpicture}[baseline=-0.5*\rd]
\draw[line width=0.5mm, blue] (-1*\rd,-1*\rd) rectangle ++(2*\rd,2*\rd);\draw[line width=0.5mm, blue] (-1*\rd,-1*\rd) rectangle ++(2*\rd,2*\rd);
\end{tikzpicture}
}

\newcommand{\ob}[1]{{\overbracket[0.5pt]{#1}}}

\begin{document}
\title{Proof of nonintegrability of the spin-$1$ bilinear-biquadratic chain model}
\author{HaRu K. Park}
\email{haru.k.park@kaist.ac.kr}
\affiliation{Department of Physics, Korea Advanced Institute of Science and Technology, Daejeon, 34141, Korea}
\author{SungBin Lee}
\email{sungbin@kaist.ac.kr}
\affiliation{Department of Physics, Korea Advanced Institute of Science and Technology, Daejeon, 34141, Korea}
\date{\today}
\begin{abstract}
Spin-$1$ chain models have been extensively studied in condensed matter physics, significantly advancing our understanding of quantum magnetism and low-dimensional systems, which exhibit unique properties compared to their spin-$1/2$ counterparts. Despite substantial progress in this area, providing a rigorous proof of nonintegrability for the bilinear-biquadratic chain model remains an open challenge. While integrable solutions are known for specific parameter values, a comprehensive understanding of the model’s general integrability has been elusive. In this paper, we present the first rigorous proof of nonintegrability for the general spin-$1$ bilinear-biquadratic chain models. Our proof not only confirms the nonintegrability of widely studied models but also extends to offer deeper insights into several areas. These include the unification of nonintegrability proofs using graph theoretical methods and the identification of the absence of local conserved quantities in quantum many-body scar systems with perfect fidelity revivals, such as the AKLT model. This work marks a significant step toward understanding the complex dynamics of spin-$1$ systems and offers a framework that can be applied to a broader class of quantum many-body systems.
\end{abstract}

\maketitle

\underline{\bf \textit{Introduction.}} --- Integrability\cite{PhysRevLett.98.050405,pozsgay2013generalized,vidmar2016generalized,doi:10.1126/science.1257026} and thermalization\cite{PhysRevA.43.2046,PhysRevE.50.888,deutsch2018eigenstate,doi:10.1126/science.aaf6725} are two fundamental and opposing concepts essential for understanding the long-time dynamics of the quantum systems. The presence of numerous local conserved quantities in integrable systems prevents thermalization, while decoherence arising from the absence of local conserved quantities in thermalizing systems leads to thermalization. However, integrability and thermalization do not represent the only possible classifications of quantum systems. A notable example of an intermediate class is a quantum many-body scar(QMBS) system,  which typically thermalizes for most low-entangled initial states but exhibits nonthermal behavior, such as short-time revivals, for a small subset of special initial states\cite{moudgalya2022quantum}. Although QMBS systems, such as the spin-$1/2$ PXP model\cite{turner2018weak} and the spin-$1$ AKLT model\cite{PhysRevB.98.235155,PhysRevB.101.195131}, are believed to lack local conserved quantities, it is thought that their dynamics are governed by low-entangled conserved quantities, which play a key role in their behavior\cite{PhysRevLett.119.030601,PhysRevX.13.031013}. Consequently, identifying conserved quantities in quantum models is crucial for gaining insight into their long-time dynamics.

\begin{table}
\centering
\begin{tblr}{l|l|l|l}
\hline
Model                               & Spin                          & Ref.   &First proof for...     \\ \hline
{$XYZ+$magnetic field}  & \SetCell[r=5]{} $\displaystyle S=\frac{1}{2}$ & \cite{shiraishi2019proof} & Spin-$1/2$\\ \hline
Mixed-field Ising                   &  & \cite{PhysRevB.109.035123}& \\ \hline
PXP                                 &  & \cite{park2024proof}& QMBS model \\ \hline
{Next-nearest\\neighbor Heisenberg} &  &\cite{Shiraishi2024}& \\ \hline
{Asymmetric\\Heisenberg-like} &  & \cite{PhysRevLett.133.170404}& \\ \hline
\SetCell[c=1]{l, green!30!}{Spin-$1$\\bilinear-biquadratic\\(including AKLT,\\Haldane chain)}    &\SetCell[c=1]{l, green!30!} $\displaystyle S=1$           &\SetCell[c=1]{l, green!30!} {This\\work}&\SetCell[c=1]{l, green!30!}{Spin-$1$ model,\\QMBS model\\with perfect revival} \\ \hline

\end{tblr}\caption{Proofs of nonintegrable models. The spin-$1$ bilinear-biquadratic model, shown in the bottom row, is the only model that has been formally proven to be nonintegrable in the spin $1$ case.}\label{table:models}
\end{table}

While various methods exist for constructing local conserved quantities in integrable models\cite{bethe1931theorie,yang1967some,baxter1978solvable,baxter2016exactly,takhtadzhan1979quantum}, rigorous approaches for nonintgrable systems are still lacking. Criteria such as level statistics\cite{PhysRevLett.54.1350} or Grabowski-Mathieu conjecture\cite{grabowski1995integrability} are often employed to diagnose nonintegrability but fall short of providing formal proofs. Only very recently have specific proofs emerged, confirming the nonintegrability of certain models in spin-1/2 systems. Table. \ref{table:models} summarizes recent studies of these models. Notably, a graph theoretical approach to prove nonintegrability with efficient time complexity has been proposed by the current authors.  
Despite a few recent studies on nonintegrability in spin-$1/2$ systems, no research has yet investigated nonintegrability in systems beyond spin-1/2. In particular, the spin-$1$ chain model with bilinear and biquadratic terms is known to be integrable only for very fine-tuned conditions\cite{lai1974lattice,sutherland1975model,barber1989spectrum,batchelor1990spin,takhtajan1982picture,babujian1982exact}. Beyond these points, the model is believed to be nonintegrable\cite{PhysRevB.101.195131,ercolessi2014analysis}, though a rigorous proof remains a challenging task.

In this paper, we provide a rigorous proof that the spin-$1$ bilinear-biquadratic model is generally nonintegrable, except in the well-known integrable cases under fine-tuned conditions. We emphasize that this is the first proof of nonintegrability for higher-spin systems beyond spin-$1/2$, with potential applications to other quantum many-body systems. Importantly, the nonintegrability of the spin-$1$ system confirms that the perfect revivals observed in the AKLT model are not governed by any local conserved quantity.

\underline{\bf \textit{Model.}} --- The spin-$1$ bilinear-biquadratic Heisenberg chain model is,  
\begin{equation}\label{eq:Ham}
H_\theta=\sum_{j=1}^{L} \left(\cos \theta(\vec{S}_{j}\cdot \vec{S}_{j+1}) + \sin \theta(\vec{S}_j \cdot \vec{S}_{j+1})^2\right),
\end{equation}
with $\theta\in (-\pi,\pi]$, under the periodic boundary condition, $\vec{S}_{L+j}=\vec{S}_{j}$. Here $\vec{S}= (X, Y, Z)$ represents the spin-$1$ operators.

\begin{figure}[t]
  \includegraphics[width=0.5\textwidth]{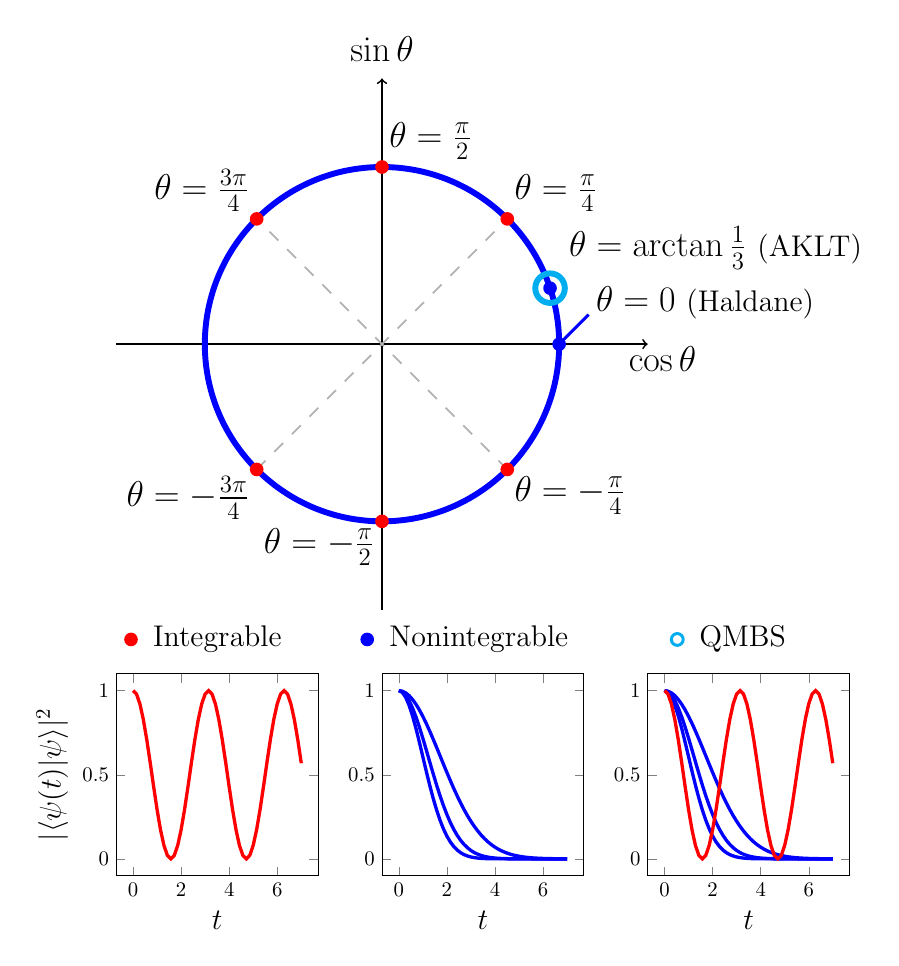}
\caption{Properties of the Hamiltonian in Eq. \ref{eq:Ham} for $\theta\in (-\pi, \pi]$. Red points mark the values of $\theta$ where the Hamiltonian is integrable, while blue regions indicate nonintegrable cases, as rigorously demonstrated in this paper. The cyan-circled point represents the AKLT model where a QMBS emerges with perfect fidelity revival. The lower panel displays the expected short-time fidelity plots for integrable, nonintegrable, and QMBS models.}
  \label{fig:integrablepoints}
\end{figure}

This model has been extensively studied for various values of $\theta$. At specific points $\theta=\pm \frac{\pi}{4},\pm\frac{3\pi}{4}, \pm\frac{\pi}{2}$, the model becomes integrable and can be analyzed by the Bethe ansatz, the Yang-Baxter equation, and the Temperley-Lieb algebra\cite{lai1974lattice,sutherland1975model,barber1989spectrum,batchelor1990spin,takhtajan1982picture,babujian1982exact}. Outside of these points, the Hamiltonian is widely believed to be nonintegrable. Notable examples with partial knowledge of certain properties include the Haldane spin-$1$ chain at $\theta=0$ and the AKLT model at $\theta=\arctan(1/3)$. These special points are highlighted in Fig. \ref{fig:integrablepoints}.

\underline{\bf \textit{Notations.}} --- Here, we introduce some notations. We define an operator $C$ as a \textit{length-$l$ operator}, denoted by $\len(C)=l$, if it can be expressed as a sum of terms acting on at most $l$ consecutive sites. For example, $\len(H_\theta)=2$. We define an operator $C$ as \textit{local} if $\len (C) < L/2 $, where $L$ is the system size.\footnote{This is reasonable because if $\len (C) \geq  L/2 $, the operator acts on a significant portion of the system and thus losing its local character.}

As a basis of $3\times 3$ matrices, define the following operators.
\begin{gather*}
\overline{X}\coloneqq YZ+ZY,\quad \overline{Y}\coloneqq ZX+XZ, \quad \overline{Z}\coloneqq XY+YX,\\
\underline{X}\coloneqq Y^2-Z^2,\quad \underline{Y}\coloneqq Z^2-X^2,\quad \underline{Z}\coloneqq X^2-Y^2.\numberthis
\end{gather*}
We denote $\mathcal{B}\coloneqq \{X,Y,Z,\allowbreak\xo,\yo,\zo,\allowbreak \xu,\yu,\zu\}$. The detailed discussion of these operators are discussed in \cite{supp}. Using the operators in $\mathcal{B}$, $H_\theta$ can be rewritten as follows:
\begin{align*}
H_\theta &=\sum_j \left(\cos\theta-\frac{\sin\theta}{2}\right)(X_jX_{j+1}+Y_j Y_{j+1}+Z_j Z_{j+1})\\
&+\sum_j \frac{\sin\theta}{3} (\xu_j\xu_{j+1} + \yu_j\yu_{j+1}+\zu_j\zu_{j+1})\\
&+\sum_j \frac{\sin\theta}{2}(\overline{X}_j \overline{X}_{j+1}+\overline{Y}_j \overline{Y}_{j+1}+\overline{Z}_j \overline{Z}_{j+1}).\numberthis\label{eq:Ham_mod}
\end{align*}
We will now use Eq. \ref{eq:Ham_mod} as the base form of the Hamiltonian for further analysis.
 
\underline{\bf \textit{Main result.}} --- \textit{For $\theta \not\in \{\pm \frac{\pi}{4},\pm\frac{3\pi}{4},\pm\frac{\pi}{2}\}$, $H_\theta$ has no local conserved quantities of length $\geq 3$.} \footnote{Here, the argument for length $1$ or $2$ is excluded which is the trivial case. It includes the Hamiltonian itself or total $z$-spin operator $\sum_{j=1}^L Z_j$.}

\underline{\bf \textit{Strategy.} }--- We first consider the absence of local conserved quantities in the presence of  translational invariance. \footnote{For the case without translational invariance, see \cite{supp}.} 
Consider the general translationally invariant quantity of length $k$:
\begin{equation}\label{eq:cons_candidate}
C=\sum_{l=1}^k \sum_{\bm{A}^l} \sum_{j=1}^L  q(\bm{A}^l) \{\bm{A}^l\}_j.
\end{equation}
Here, $\{\bm{A}^l\}_j$ represents operator strings of length $l$ that do not start or end with the identity operator $I$; the first operator acts at position $j$, the second at position $j+1$, and so on. The coefficients $q(\bm{A}^l)$ are associated with these operator strings. Since both $C$ and $H_\theta$ are translationally invariant, the commutator, $[C,H_\theta]$, is also translationally invariant and can be expressed as follows:
\begin{equation}\label{eq:comm_rel}
[C,H_\theta]=\sum_{l=1}^{k+1} \sum_{\bm{B}^l} \sum_{j=1}^L p(\bm{B}^l) \{\bm{B}^l\}_j,
\end{equation}
where $\{\bm{B}^l\}_j$ and $p(\bm{B}^l)$ are again operator strings of length $l$ and their coefficients, respectively. Substituting Eq.\eqref{eq:cons_candidate} in Eq.\eqref{eq:comm_rel} results in a series of linear equations with $q(\bm{A}^l)$ and $p(\bm{B}^l)$ as parameters. The local conserved quantity condition $[C,H]=0$ requires $p(\bm{B}^l)=0$ for all $\bm{B}^l$.

To show the absence of a length $k$ local conserved quantity in $H_\theta$, it suffices to show that $p(\bm{B}^l)=0$ for all $\bm{B}^l$ with $1\leq l\leq k+1$ in Eq.\eqref{eq:comm_rel} implies $q(\bm{A}^k)=0$ for all $\bm{A}^k$ in Eq.\eqref{eq:cons_candidate}, i.e. $C|_k=0$, where $C|_k$ is the sum of length $k$ operator strings in $C$. Now, consider the commutator $[\{\bm{A}^k\}_j,H_\theta]$. Since every operator string in $H_\theta$ is of length $2$, each operator string in $[\{\bm{A}^k\}_j,H_\theta]$ must have length $l-1$, $l$, or $l+1$. Consequently, $q(\bm{A}^l)$'s are only related to $p(\bm{B}^l)$ and $p(\bm{B}^{l \pm 1})$.

\begin{figure}[h!]
  \includegraphics[width=0.5\textwidth]{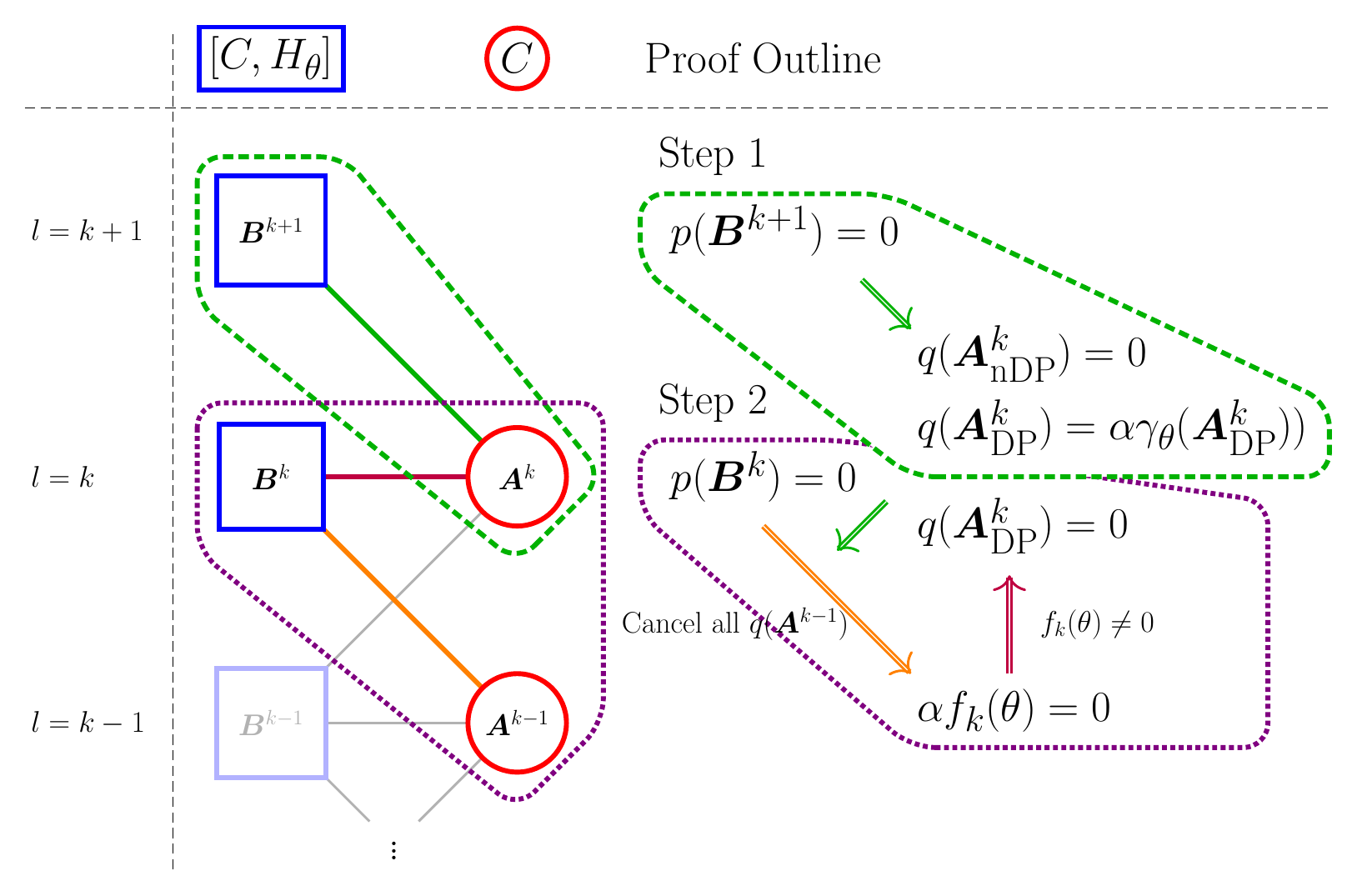}
\caption{Structure of our proof. Red circles represent the operator strings in $C$, and blue squares represent the operator strings in $[C,H_\theta]$. These operator strings are grouped by their length. The edges connecting red circles and blue squares indicate how operator strings in $C$ contribute to those in $[C,H_\theta]$. For example, the red circle labeled $\bm{A}^{k-1}$ is connected to the blue squares labeled $\bm{B}^{k-2}, \bm{B}^{k-1},$ and $\bm{B}^{k}$, showing that length $k$ operator strings in $C$ contribute to length $k-2, k-1,$ and $k$ operator strings in $[C,H_\theta]$. The double-lined arrows on the right side represent the logical flow in Steps 1 and 2, with each arrow corresponding to an edge of the same color, indicating its role in the respective logical step. The operators $\bm{A}^k_{\text{nDP}}$ and $\bm{A}^k_{\text{DP}}$ are length $k$ operator strings whose coefficients are shown to be zero in Steps 1 and 2, respectively.}
  \label{fig:proofstructure}
\end{figure}

Based on this observation, the proof outline is as follows, with a schematic diagram shown on the right-hand side of Fig.\ref{fig:proofstructure}. In Step 1, suppose a length $k$ operator $C$ satisfies $[C,H_\theta]=0$. Then, by deriving the relation between $p(\bm{B}^{k+1})=0$ and $q(\bm{A}^k)$, we show that many operator strings of length $k$ in $C$ must have zero coefficients, except for certain special operator strings, so called \textit{doubling product strings} which will be discussed later. The coefficients of these doubling product strings are uniquely determined as $\alpha \gamma_\theta$, where $\alpha$ is some constant and $\gamma_\theta$ is a function of $\theta$ that is defined as Eq.\ref{eq:appropriate_coeff} for given string operators. Our first step is to define and characterize these special operator strings.

In Step 2, we proceed as follows:  (i) Select the appropriate $\bm{B}^k_t$ operator strings defined in Eq.\ref{eq:bt1} and \ref{eq:bt2}, and focus on a series of linear equations derived from the condition $p(\bm{B}^k_t)=0$, which relates the coefficients $q(\bm{A}^k)$ and $q(\bm{A}^{k-1})$. Next, (ii) eliminate all terms involving $q(\bm{A}^{k-1})$, and substitute $q(\bm{A}^k)$ with $\gamma_\theta$, as obtained in Step 1, to arrive at a single $\theta$-dependent equation, namely $\alpha f_k(\theta)=0$. From this equation, it follows that if a length $k$ conserved quantity exists for $H_{\theta}$, then $\alpha f_k(\theta)=0$. We will show that (iii) $f_k(\theta)\neq 0$ for all $k\geq 3$ when $\theta \not\in  \{\pm \frac{\pi}{4},\pm\frac{3\pi}{4},\pm\frac{\pi}{2}\}$, which implies that if a length $k$ operator $C$ satisfies $[C,H]=0$, then $\alpha=0$. Consequently, this leads to $q(\bm{A}^k_{\text{DP}})=0$  and thus $C|_k = 0$. Therefore, we conclude that $H_\theta$ is nonintegrable for these values of $\theta$.

\begin{figure}[h!]
  \includegraphics[width=0.4\textwidth]{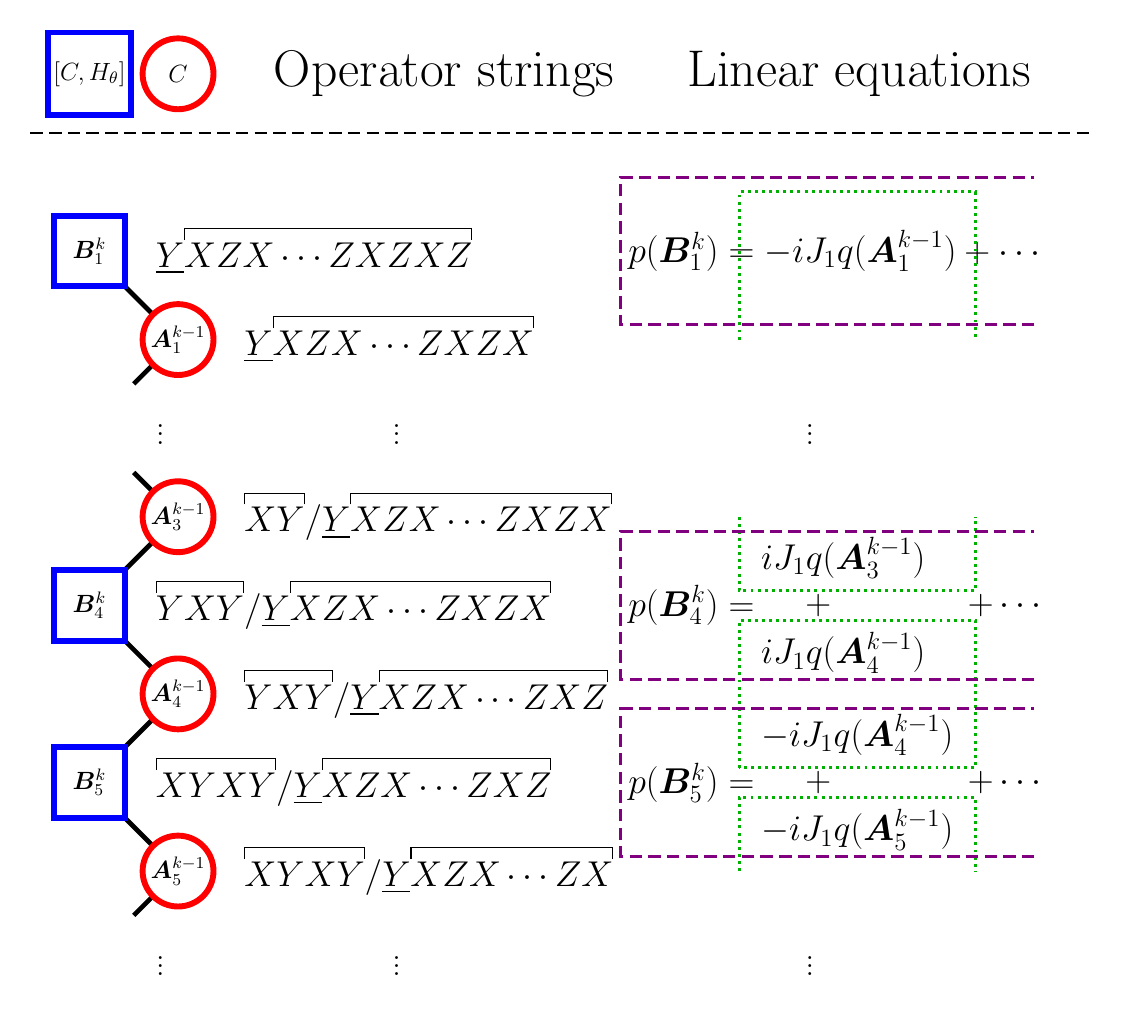}
\caption{Graph-theoretical representation of the proof in Step 2, illustrating the cancellation of $q(\bm{A}^{k-1})$ coefficients for even $k$. The blue squares $\protect\bsv$ represent the operators $\bm{B}_t^k$ in the commutator $[C,H_\theta]$, while the red circles $\protect\rcv$ represent the operators $\bm{A}_t^{k-1}$ in $C$. The explicit form of each operator is shown next to its respective $\protect\rcv$/$\protect\bsv$ symbol. A $\protect \rcv$ vertex $\bm{A}_t^{k-1}$ is connected to a $\protect \bsv$ vertex $\bm{B}_{t'}^{k}$ by an edge when $\bm{A}_t^{k-1}$ contributes nontrivially to $\bm{B}_{t'}^{k}$. The linear equations corresponding to each $\bm{B}_t^k$ are presented in the purple boxes on the right, showing only the contributions from $\bm{A}^{k-1}$ operators, while omitting the $q(\bm{A}^k)$ terms. The coefficients in the green box cancel each other, allowing for the elimination of all $q(\bm{A}_t^{k-1})$ coefficients via substitution.}
  \label{fig:step2_short}
\end{figure}

The left-hand side of Fig.\ref{fig:proofstructure} shows a simplified sketch of the graph-theoretical approach with blue squares and red circles, which is described in more detail in Fig.\ref{fig:step2_short}.

\underline{\bf \textit{Nonintegrability for $\bm{\theta \not\in\{\pm \frac{\pi}{4},\pm\frac{3\pi}{4},\pm\frac{\pi}{2}\}}$.} }---
We now present the exact statements for each step outlined in the previous section, along with the main ideas behind their proofs. 

{\bf \textit{Step 1.}} --- In the first step, we set $p(\bm{B}^{k+1})= 0$ and show that only specific operator strings, called \textit{doubling product strings} and denoted as $\bm{A}^k_{\text{DP}}$, can have non-zero coefficients. All other operator strings, denoted by $\bm{A}^k_{\text{nDP}}$, must have zero coefficients. A doubling product string is defined as an operator string that has a non-zero coefficient in the following \textit{doubling product operator}: 
\begin{align*}\label{eq:dpform}
&\{\ob{A_{(1)}A_{(2)}\cdots A_{(n)}}\}_j \\
&= c \cdot (A_{(1)})_j ([A_{(1)}, A_{(2)}])_{j+1} ([A_{(2)}, A_{(3)}])_{j+2} \cdots\\
&\cdots ([A_{(n-1)},A_{(n)}])_{j+n-1} (A_{(n)})_{j+n}, \numberthis
\end{align*}
where, $(A_{(t)})_j$ denotes a single-site operator $A_{(t)}\in \mathcal{B}$ acting at site $j$, and $c$ is a complex number chosen to normalize the coefficient to $1$. Note that different doubling product forms can yield the same doubling product operator. For example, $\ob{X\xu Y} = \ob{X\yu Y} = X\xo\yo Y$.

Now we define a function $\gamma_\theta$ as:
\begin{equation}
\gamma_\theta(\bm{D}) \!=\! \!\!  \! \!\!  \sum_{\bm{D}=\ob{\scriptstyle \bm{A}^{(k-1)}}} \prod_{n=1}^{k-2} \coeff ([A_{(n)},A_{(n+1)}])\prod_{n=1}^{k-1} J_\theta (A_{(n)}). \label{eq:appropriate_coeff}
\end{equation}
Here, $\bm{D}$ is a doubling product operator, defined with $\bm{A}^{(k-1)}=A_{(1)}\cdots A_{(k-1)}$. $\coeff([A{(j)},A_{(j+1)}])$ represents the coefficient of commutator divided by $i$(for example, $\coeff ([X,\xu])=2$), and $J_\theta (A)$ is defined as follows:
\begin{equation}
J_\theta (A)=\begin{cases}
\cos\theta-\frac{1}{2}\sin\theta&A=X, Y, Z\\
\frac{1}{2}\sin\theta&A=\xo, \yo,\zo\\
\frac{1}{3}\sin\theta&A=\xu,\yu,\zu
\end{cases}
\end{equation}
For example, $\gamma_\theta(XZY) = (\cos\theta-\sin\theta/2)^2$ and $\gamma_\theta(Y\yo\xo X) = -(\cos\theta-\sin\theta/2)^2\sin\theta$. Note that $J_\theta (A)$ corresponds to the coefficient of $AA$ in Eq. \ref{eq:Ham_mod}.

{\bf \textit{Statement.}}
\textit{Let $C$ be a length $k$ local quantity satisfying $\len([C,H_\theta])\leq k$, i.e., $p(\bm{B}^{k+1})=0$. Then 
\begin{equation}
C|_k=\alpha\sum_{\bm{D}, j} \gamma_\theta (\bm{D})\cdot \{\bm{D}\}_j, \label{eq:step1}
\end{equation}
where the sum runs over the set of doubling product operators, $\alpha$ is an overall factor of length $k$ operator strings and $C|_k$ is the sum of length $k$ operators in $C$.} This statement asserts that if $\len([C,H_\theta])\leq k$ then the coefficients of $\bm{A}^k_{\text{DP}}$ are fully determined by the function $\gamma_\theta $, and the coefficients of $\bm{A}^k_{\text{nDP}}$ must be $0$.

{\bf \textit{Proof.} } We provide a brief sketch of the proof here; see \cite{supp} for full details. Although not every operator string is a doubling product string, we can always identify a substring on the left that is part of a doubling product string. Using the condition $\len[C,H_\theta]\leq k$, for a non-doubling operator string $\bm{A}$, one can find that $q(\bm{A})$ is proportional to $q(\bm{A}')$, where the leftmost substring of $\bm{A}'$ that can be represented as a doubling product is shorter than the one for $\bm{A}$; if this substring cannot be shortened further, then $q(\bm{A})=0$. Thus, by induction, we can show that for any non-doubling product string $\bm{A}$, $q(\bm{A})=0$.

Using a similar argument, for a doubling operator string $\bm{A}$, one finds that $q(\bm{A})$ is proportional to $q(\bm{A}')$, where $\bm{A}'$ is another doubling operator string. This implies that if the coefficient of one doubling operator string is fixed, all other coefficients are correspondingly determined; in other words, $C|_k$ is uniquely determined up to a scaling factor.

Let $H_\theta =\sum_j h_j$, where $h_j$ is an operator acting only on sites $j$ and $j+1$. Define,
\begin{equation}
C_{\text{boost}}\coloneqq \sum_j [h_{j},[h_{j+1},[\cdots ,[h_{j+k-3},h_{j+k-2}]\cdots]]].
\end{equation}
It can be shown that $\len [C_{\text{boost}},H_\theta]\leq k$. Since $C|_k$ is uniquely determined up to a scaling factor, we have $C|_k = \alpha C_{\text{boost}}$ for some constant $\alpha$. Finally, it can be proven that $C_{\text{boost}}=\sum_{\textbf{D},j}\gamma_\theta(\bm{D})\cdot \{\bm{D}\}_j$, thus proving the statement. $\square$

{\bf \textit{Step 2.} }--- In Step 1, we have shown all $q(\bm{A}^k_{\text{nDP}})=0$ under the condition $p(\bm{B}^{k+1})=0$, and $q(\bm{A}^k_{\text{DP}})$ is determined by a single parameter $\alpha$. Now we prove that \textit{if $p(\bm{B}^k)=0$ and $\theta \not\in\{\pm \frac{\pi}{4},\pm\frac{3\pi}{4},\pm\frac{\pi}{2}\}$, then $\alpha=0$ and thus every $q(\bm{A}^k_{\text{DP}})$ is zero.}


To simplify notation for the appropriate $\bm{B}^k$-type operator strings, we introduce a slash notation between two operator strings of lengths $n$ and $m$, respectively, defined as follows:
\begin{align*}
&A_{(1)}\cdots A_{(n-1)}A_{(n)}/B_{(1)}B_{(2)}\cdots B_{(m)} \\
&=c\cdot  A_{(1)}\cdots A_{(n-1)}([A_{(n)},B_{(1)}])B_{(2)}\cdots B_{(m)}\numberthis
\end{align*}
This is an operator string of length $n+m-1$
Here, $c$ is a complex number chosen to normalize the coefficient to $1$. For example, $XY/ZZ = XXZ$. Also, for an operator string $\bm{A}$, let $(\bm{A})^{(k)}$ denote a $k$-fold repetition of $\bm{A}$ operator. For instance, $(XZ)^{(2)} = XZXZ$.

Define the following operator strings for even and odd values of $k$:
\begin{equation}\label{eq:bt1}
\bm{B}_{t}^{2m} \coloneqq \begin{cases}
\ob{(XY)^{(n)}}/\yu\ob{(XZ)^{(m-n-1)}},&t=2n+1\\
\ob{Y(XY)^{(n-1)}}/\yu  \ob{(XZ)^{(m-n-1)}X},&t=2n \\
\end{cases}
\end{equation}
\begin{equation}\label{eq:bt2}
\bm{B}_{t}^{2m+1} \coloneqq \begin{cases}
\ob{(XY)^{(n)}}/\yu\ob{(XZ)^{(m-n-1)}X},&t=2n+1\\
\ob{Y(XY)^{(n-1)}}/\yu  \ob{(XZ)^{(m-n)}},&t=2n 
\end{cases}
\end{equation}
Here, compared to the doubling product form, there exists a discrepancy $\ob{\cdots Y}/\yu\ob{X\cdots}$ in the middle of each $\bm{B}_t^k$. The key point for the proof of Step 2. lies on that every operator string in $C$ that contribute to each $\bm{B}_t^k$ must be either (i) be a length $k-1$ operator string or (ii) be a a length $k$ operator string represented as doubling product operator.

First, consider the operator strings of length $k-1$ that contribute to $\bm{B}_t^k$. Due to the construction of $\bm{B}_t^k$, at most two different length $k-1$ operator strings contribute to each $\bm{B}_t^k$. For example, for odd $t$ and even $k$, $\bm{B}_t^k$ satisfies:
\begin{align*}
&\ob{XY\cdots XY}/\yu\ob{XZ\cdots XZ} \\
=&\tcr{\ob{X}}/(\ob{Y\cdots XY}/\yu\ob{XZ\cdots XZ})\\
=&(\ob{XY\cdots XY}/\yu\ob{XZ\cdots X})/\tcr{\ob{Z}}.\numberthis
\end{align*}
Here the red-colored operator strings $\ob{X}$ and $\ob{Z}$ come from the Hamiltonian $H_\theta$, while the operator strings next to them are length $k-1$ operator strings from $C$. For $t=1$ or $t=k-2$, only a single operator string of length $k-1$ contributes to $\bm{B}_t^k$:
\begin{align*}
\yu \ob{XZ\cdots XZ} &= \yu \ob{XZ\cdots X}/\tcr{\ob{Z}}\\
\ob{YX\cdots YXY}/\yu \ob{X} &= \tcr{\ob{Y}}/\ob{X\cdots YXY}/\yu \ob{X}.\numberthis
\end{align*}
Finally, $\bm{B}_t^k$ and $\bm{B}_{t+1}^k$ share a single length $k-1$ operator strings, denoted $\bm{A}_t^{k-1}$, which contributes to both. The left panel of Fig.\ref{fig:step2_short} illustrates this situation graphically for even $k$.\footnote{The graph structure in Fig.\ref{fig:step2_short} is also a key element in proving nonintegrability for spin-$1/2$ systems, using the graph-theoretical framework\cite{park2024proof}.}

After writing the series of linear equations corresponding to each $\bm{B}_t^k$ operator string and using traditional substitution method, we can remove the variables $q(\bm{A}_t^{k-1})$ from the system of equations. The right panel of Fig.\ref{fig:step2_short} illustrates this cancellation of coefficients. Therefore, after setting $p(\bm{B}_t^k)=0$, we arrive at a single equation composed solely of the coefficients $q(\bm{A}_{\text{DP}}^k)$.\footnote{Due to the linear dependence of $\xu, \yu,$ and $\zu$, an additional set of equations is required to fully eliminate the coefficients of length $k-1$ operator strings. For more details, see \cite{supp}. However, the main idea remains unaffected by this slight modification.}

Since Eq.\ref{eq:step1} shows that the coefficients $q(\bm{A}^k_{\text{DP}})$ depend on a single parameter $\alpha$, after substitution, the resulting equation takes the form $\alpha f_k(\theta) = 0.$ Thus, our final goal is to explicitly write $f_k (\theta)$ to complete the proof. Although many commutators contributes to the function $f_k(\theta)$, by leveraging the locality of the Hamiltonian $H_\theta$, the calculation simplifies, yielding the exact form of $f_k(\theta)$ as follows:
\begin{equation}\label{eq:fk}
f_k(\theta) =  (k-1) \cos\theta \cos 2\theta.
\end{equation}
If  $\cos\theta \cos 2\theta \neq 0$, we find that $f_k(\theta)\neq 0$ and therefore $\alpha=0$, which implies that $q(\bm{A}^k_{\text{DP}})=0$ for all doubling product operators. The condition $\cos\theta \cos 2\theta \neq 0$ holds when $\theta\not\in \{\pm \frac{\pi}{4}, \pm\frac{3\pi}{4},\pm\frac{\pi}{2}\}$, and it is known that $\theta\in \{\pm \frac{\pi}{4}, \pm\frac{3\pi}{4},\pm\frac{\pi}{2}\}$ corresponds to the integrable cases of $H_\theta$. This completely classifies the integrability and non-integrability of $H_\theta$. A similar analysis also applies to $\xo, \yo, \zo$ by substituting $X,Y,Z$ in Eq.\ref{eq:bt1} and Eq.\ref{eq:bt2}; for more details, see \cite{supp}.

\textit{Discussion.} --- We have rigorously proven that the spin-$1$ bilinear-biquadratic chain model lacks local conserved quantities, establishing its general nonintegrability, except at known integrable points. Our findings eliminates the possibility of hidden local symmetries in prominent spin-$1$ chain models, such as the Haldane spin-$1$ chain and the AKLT model. Notably, our proof shows the nonintegrable nature of the AKLT model, marking a significant result as it provides the first rigorous evidence of nonintegrability in quantum many-body scarring systems exhibiting perfect fidelity revivals. It also shows that local conserved quantities are not essential for the emergence of unique physical phenomena, such as topologically protected gaps or short-time fidelity revivals.


Our study extends beyond previous nonintegrability research in spin-$1/2$ systems, addressing the unique challenges posed by spin-$1$ systems' complex operator commutation relations. The complexity of higher-spin systems has traditionally posed challenges to proving nonintegrability. This study demonstrates that nonintegrability can instead be established based on Hamiltonian locality rather than solely on single-site commutation relations, indicating potential applicability to a broader range of local Hamiltonians.
In addition, we believe that our approach can be generalized to prove nonintegrability in higher-spin Heisenberg-like models. For any spin-$S>1/2$, the quadratic spin operators $X^2, Y^2, Z^2$ remain linearly dependent, satisfying $X^2+Y^2+Z^2 = I_{2S+1}$, but are not proportional to the identity operator. This suggests that similar argument could play a crucial role in the proof of nonintegrability for higher-spin systems, similar to the role it plays in the spin-$1$ model.

Furthermore, since we have shown that the Hilbert space dimension of the single-site operator does not affect the standard strategy for proving nonintegrability, we believe our approach could provide insight into the Grabowski-Mathieu conjecture, which claims that the absence of length $3$ conserved quantities implies nonintegrability of the system. We leave this extension to higher-spin systems as future work.


\begin{acknowledgments}

\noindent	
{\em Acknowledgments.---}
This work is supported by National Research Foundation (NRF) Grant No. 2021R1A2C1093060 and Nano Material Technology Development Program through the NRF funded by Ministry of Science and ICT (RS-2023-00281839).
\end{acknowledgments}

\bibliography{biblio}

\end{document}


\title{Proof of nonintegrability of the spin-$1$ bilinear-biquadratic chain mode:\\ Supplement Material}
\author{HaRu K. Park, SungBin Lee}
\maketitle

\tableofcontents

\section{Single-Site Operators and Completeness of the Basis}

In the main text, we define $\mathcal{B}$ as the set of single-site operators:
\begin{equation}
\mathcal{B} =\{ X,Y,Z, \xo, \yo, \zo, \xu, \yu, \zu\},
\end{equation}
where the operators are specified as follows:
\begin{gather}
X\coloneqq \frac{1}{\sqrt{2}}\begin{bmatrix}
0&1&0\\
1&0&1\\
0&1&0
\end{bmatrix},\quad Y\coloneqq \frac{1}{\sqrt{2}}\begin{bmatrix}
0&-i&0\\
i&0&-i\\
0&i&0
\end{bmatrix},\quad
Z\coloneqq \begin{bmatrix}
1&0&0\\
0&0&0\\
0&0&-1
\end{bmatrix}\\
\xo \coloneqq YZ+ZY,\quad \yo\coloneqq ZX+XZ,\quad \zo\coloneqq XY+YX\\
\xu=Y^2-Z^2,\quad \yu=Z^2-X^2,\quad \zu=X^2-Y^2.
\end{gather}
While nine operators are necessary for a complete basis of Hermitian $3\times 3$ matrices, $\mathcal{B}$ is not itself a basis  due to the linear dependence $\xu+\yu+\zu=0$. The following set of ten single-site operators forms an overcomplete basis:
\begin{equation}
\mathcal{B}_I = \{ X,Y,Z, \xo, \yo, \zo, \xu, \yu, \zu, I\}.
\end{equation}
To eliminate this overcompleteness, we select the following set as our basis for Hermitian $3\times 3$ matrices:
\begin{equation}
\mathcal{B}_0 = \{X,Y,Z,\xo,\yo,\zo,\xu,\yu,I\}.
\end{equation}
Although $\mathcal{B}_0$ serves as our primary basis, we occasionally use other symbols, such as $\zu$, when it simplifies the arguments. However, we rely on $\mathcal{B}_0$ for completeness when needed.

The completeness of $\mathcal{B}_0$ is demonstrated by the following expression:
\begin{align*}
\begin{bmatrix}
u&a+ib&c+id\\
a-ib&v&e+if\\
c-id&e-if&w
\end{bmatrix}&=u\frac{3Z+2I-\underline{X}+\underline{Y}}{6}\\
&+v\frac{I-\underline{X}-\underline{Y}}{3}\\
&+w\frac{-3Z+2I-\underline{X}+\underline{Y}}{6}\\
&+a\frac{X+\overline{Y}}{\sqrt{2}}-b\frac{Y+\overline{X}}{\sqrt{2}}\\
&+e\frac{X-\overline{Y}}{\sqrt{2}}-f\frac{Y-\overline{X}}{\sqrt{2}}\\
&-c(\underline{X}+\underline{Y})-d\overline{Z}
\end{align*}

The multiplication and commutation relations for the single-site operators are provided in Section \ref{section:7}.

Under the basis $\mathcal{B}_0$, we express our Hamiltonian as follows:
\begin{align*}
H_\theta &=\sum_j \left(\cos\theta-\frac{\sin\theta}{2}\right)(X_jX_{j+1}+Y_j Y_{j+1}+Z_j Z_{j+1})\\
&+\sum_j \frac{\sin\theta}{3} (2\xu_j\xu_{j+1} + 2\yu_j\yu_{j+1}+\xu_j\yu_{j+1}+\yu_j\xu_{j+1})\\
&+\sum_j \frac{\sin\theta}{2}(\overline{X}_j \overline{X}_{j+1}+\overline{Y}_j \overline{Y}_{j+1}+\overline{Z}_j \overline{Z}_{j+1}).\numberthis\label{eq:Ham_modifiedform}
\end{align*}
We frequently employ this form of the Hamiltonian to verify completeness. However, for convenience in calculating the doubling product operators, we also use the Hamiltonian form presented in the main text:
\begin{align*}
H_\theta &=\sum_j \left(\cos\theta-\frac{\sin\theta}{2}\right)(X_jX_{j+1}+Y_j Y_{j+1}+Z_j Z_{j+1})\\
&+\sum_j \frac{\sin\theta}{3} (\xu_j\xu_{j+1} + \yu_j\yu_{j+1}+\zu_j\zu_{j+1})\\
&+\sum_j \frac{\sin\theta}{2}(\overline{X}_j \overline{X}_{j+1}+\overline{Y}_j \overline{Y}_{j+1}+\overline{Z}_j \overline{Z}_{j+1}).\numberthis\label{eq:Ham_mod}
\end{align*}

\section{Doubling Product operator}
Before going through the detailed proof, we need to precisely define the concept of a doubling product operator and discuss its properties.

A doubling product operator is an operator expressed in the following doubling product form:
\begin{align*}\label{eq:dpform}
\{\ob{A_{(1)}A_{(2)}\cdots A_{(n)}}\}_j =& c \cdot (A_{(1)})_j ([A_{(1)}, A_{(2)}])_{j+1} ([A_{(2)}, A_{(3)}])_{j+2} \cdots\\
&\cdots ([A_{(n-1)},A_{(n)}])_{j+n-1} (A_{(n)})_{j+n}, \numberthis
\end{align*}
where $c$ is a normalization constant, and each $A_{(t)}$ belongs to the set $\mathcal{B}$.

It is important to note that different doubling product forms can yield the same operator. For example, $\ob{X\xu Y} = \ob{X\yu Y}=X\xo\yo Y$. This redundancy arises from the properties of $\xu, \yu,$ and $\zu$. As shown in Table \ref{table:comm}, the commutators $[\xu,A], [\yu,A], [\zu,A]$ always yield the same operator up to a scaling factor. This property introduces flexibility in the notation for doubling product operators, allowing interchangeable use of $\xu, \yu,$ and $\zu$. While this arbitrariness could be restricted by allowing only $\xu$ for $A_{(t)}$ and excluding $\yu$ and $\zu$, we will later see that this flexibility helps simplify the conditions for $C|_{\len(C)}$ in Step 1.

Furthermore, a doubling product operator does not always correspond to a single element in the basis of the operator strings, which are constructed from the single-site operators $\mathcal{B}_0$. For example, $\ob{Z\zo} = Z\zu \zo=-Z\xu \zo - Z\yu \zo$. Thus, while the operator string $Z\xu \zo$ does not directly appear in doubling product form, it is part of a doubling product form and, therefore, cannot be assumed to have a zero coefficient in Step 1 of the main text. To formalize this, we denote $\bm{A} \leq \bm{B}$ for an operator string $\bm{A}$ and operator $\bm{B}$ when $\bm{A}$ has a nonzero coefficient in the basis representation of $\bm{B}$. For example, $Z\xu\zo \leq \ob{Z\zo}$ and $Z\yu\zo\leq \ob{Z\zo}$.

A natural question arises: how can we determine whether a specific operator string $\bm{A}=A_{(1)}A_{(2)}\cdots A_{(n)}$ can be expressed in a doubling product form $\ob{D_{(1)}\cdots D_{(n-1)}}$, or at least as part of a doubling product form? Recalling the construction of the doubling product operators, this can be achieved through the following steps:
\begin{enumerate}
\item First, set $D_{(1)}\in \mathcal{B}$ to $A_{(1)}$.
\item Next, choose an operator $D_{(2)}\in \mathcal{B}$ that satisfies $A_{(2)}\leq [D_{(1)},D_{(2)}]$.
\item Proceed inductively, selecting $D_{(j)}$ such that $A_{(j)}\leq [D_{(j-1)},D_{(j)}]$, continuing this process until $j=n-1$.
\item After completing these inductive steps, if the final single-site operator $D_{(n-1)}$ satisfies $A_{(n)}\leq D_{(n-1)}$, then we have successfully found a doubling product operator such that $\bm{A}\leq \ob{D_{(1)}\cdots D_{(n-1)}}$. If this condition is not satisfied, $\bm{A}$ cannot be represented as (or included in) a doubling product form.
\end{enumerate}

In this process, the crucial point is as follows: \textit{For given single-site operators $B$ and $F$, with $B\in \mathcal{B}_0$ and $F\in \mathcal{B}$, can we always find an operator $R\in \mathcal{B}$ such that $B\leq [R,F]$?} Naturally, the answer depends on the specific operators $B$ and $F$. For example, if $(B,F)=(X,Y)$ then one can find $R=Z$, whereas if $(B,F)=(X,X)$, no such $R$ exists. Crucially, whether such an $R$ can be found is determined solely by the commutator properties of the single-site operators, without requiring any additional information, such as the commutator relations between two-site operators.

\begin{figure}[t] 
\centering
\includegraphics[width=1\linewidth]{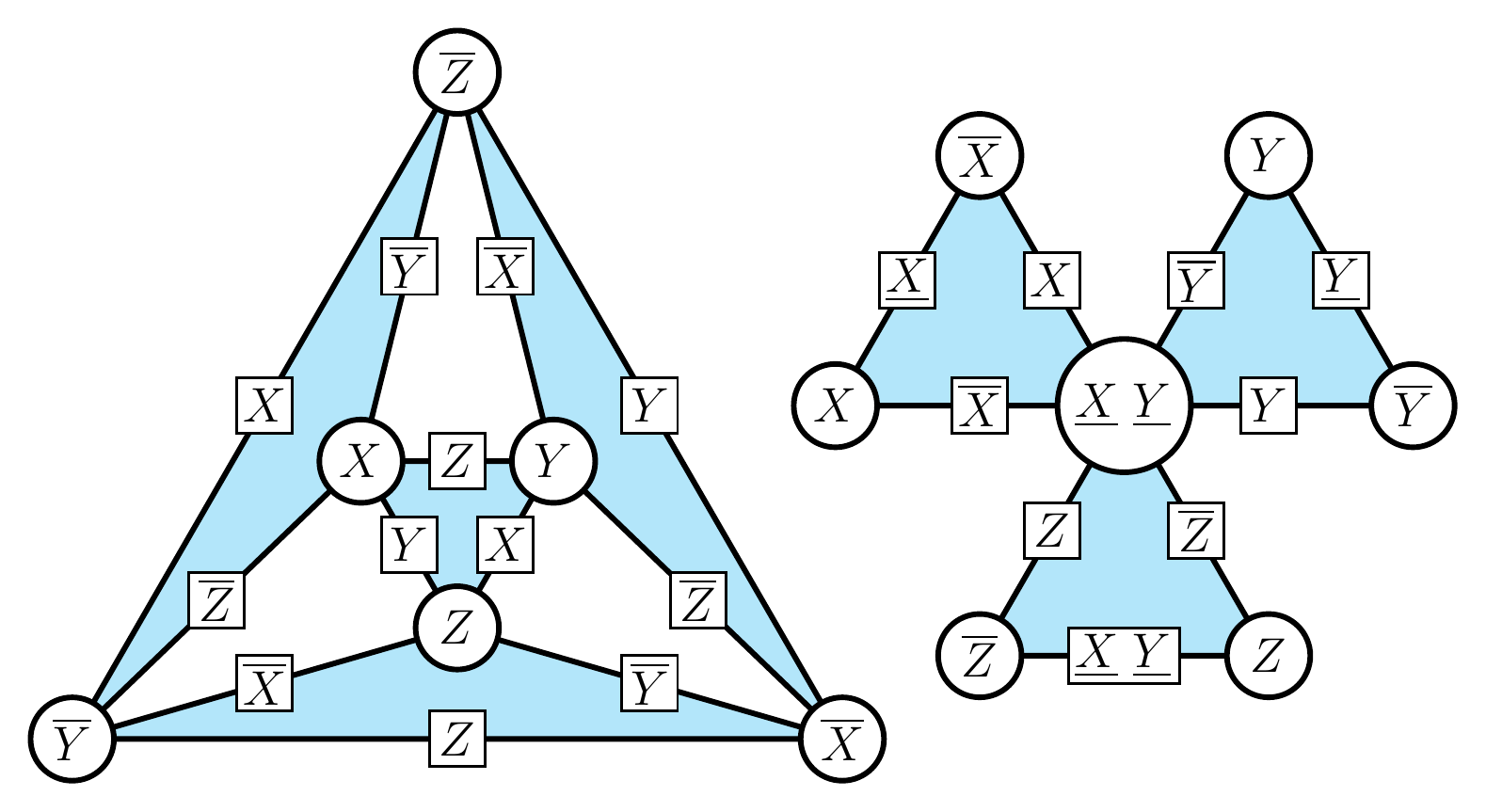}
\caption{Graphical representation of the commutation relations between single-site operators as listed in Table \ref{table:comm}. The commutator of two circled operators is represented by the basis elements within the squared operators positioned between them. If two circled operators are not connected, they commute. Some vertices display the same operator symbol multiple times for clarity; these vertices represent identical operators.}
  \label{fig:comm_structure} 
\end{figure}

Fig.\ref{fig:comm_structure} illustrates the commutation relations between single-site operators. If two operators $A_1,A_2$ within a circle are connected by an edge labeled with an operator $B$, then $B\leq [A_1,A_2]$. Notably, since $[\zo,Z]= 2i\zu = -2i\xu - 2i\yu$, there are two operators, $\xu$ and $\yu$, positioned between $\zo$ and $Z$.

This graphical representation suggests a straightforward strategy for determining whether an operator string $\bm{A}$ satisfies $\bm{A}\leq \ob{D_{(1)}\cdots D_{(n-1)}}$: 
\begin{enumerate}
\item First, select a circle that contains the first operator in $\bm{A}$, denoted $A_{(1)}$.
\item Next, choose an edge that starts from this circle and includes the second operator in $\bm{A}$, $A_{(2)}$.
\item Repeat this process inductively, finding a path with $n-1$ circled vertices and $n-2$ edges.
\item After these steps, if the final vertex reached contains the last operator of $\bm{A}$, $A_{(n)}$, then the process is complete; assigning $D_{(j)}$ as the operator in $j$-th circled vertex yields $\bm{A}\leq \ob{D_{(1)}\cdots D_{(n-1)}}$. If this process fails, e.g., no suitable edge is available in steps 2 or 3, or the last vertex does not contain the final operator, then no such doubling product operator exists.
\end{enumerate}
Each step in this strategy aligns directly with the previous approach. 

\begin{figure}[t] 
\centering
\includegraphics[width=1\linewidth]{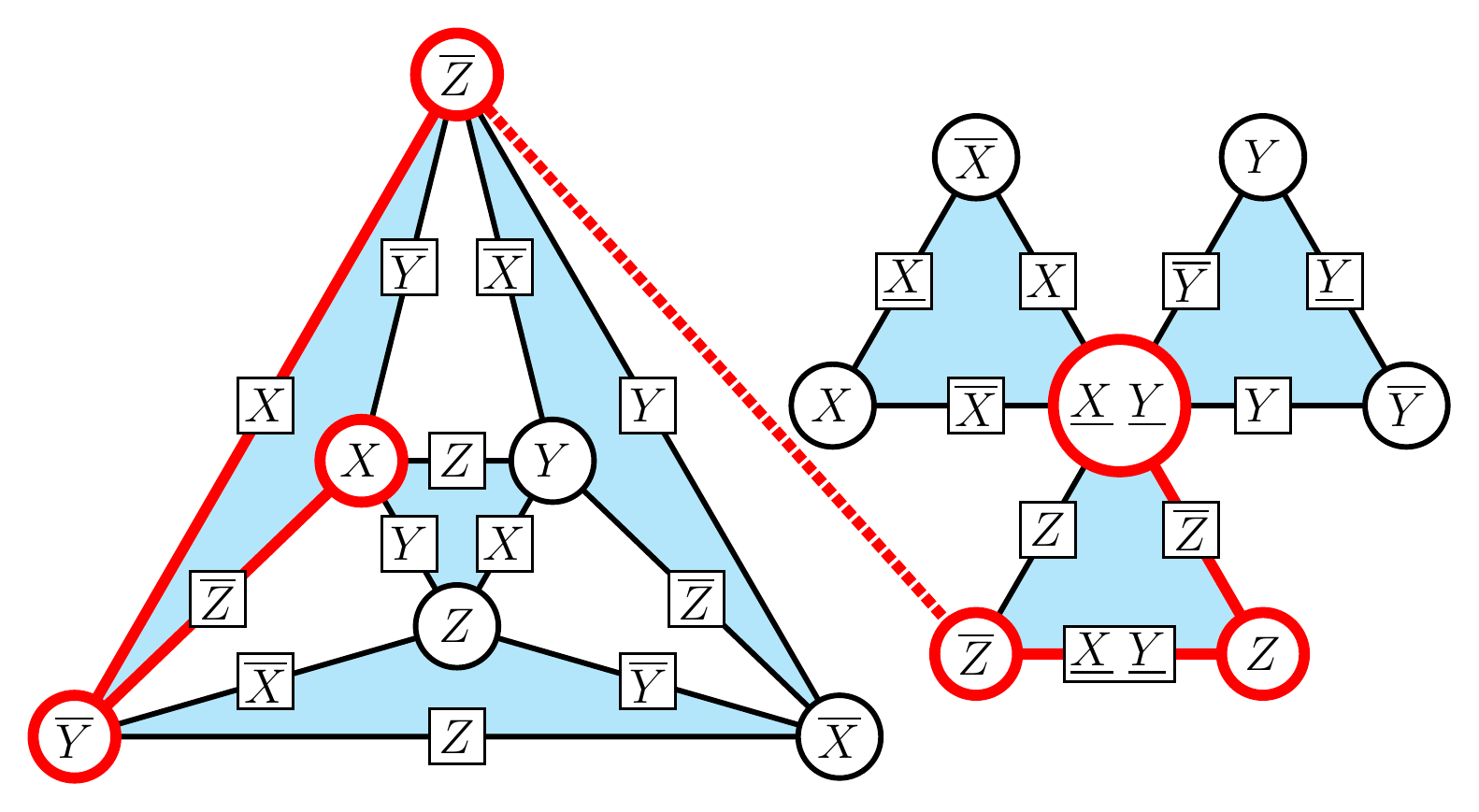}
\caption{Path used to obtain the relation $X\zo X\yu \zo \yu \leq \protect\ob{X\yo\zo Z \yu}$, as derived by our strategy. The solid line indicates the selected path, while the dotted line connects pairs of identical operators.}
  \label{fig:comm_structure_pathfind} 
\end{figure}

Fig.\ref{fig:comm_structure_pathfind} illustrates this strategy applied to the operator string $\bm{A}=X\zo X\yu\zo\yu$, resulting in the relation $\bm{A}\leq \ob{X\yo\zo Z\yu}.$ Note that this strategy is deterministic: it always yields the same doubling product operator for a given operator string, except for the option to interchange $\xu$ and $\yu$. While this substitution changes the \textit{form} of the doubling product, the doubling product \textit{operator} remains same.

\begin{figure}[t] 
\centering
\includegraphics[width=1\linewidth]{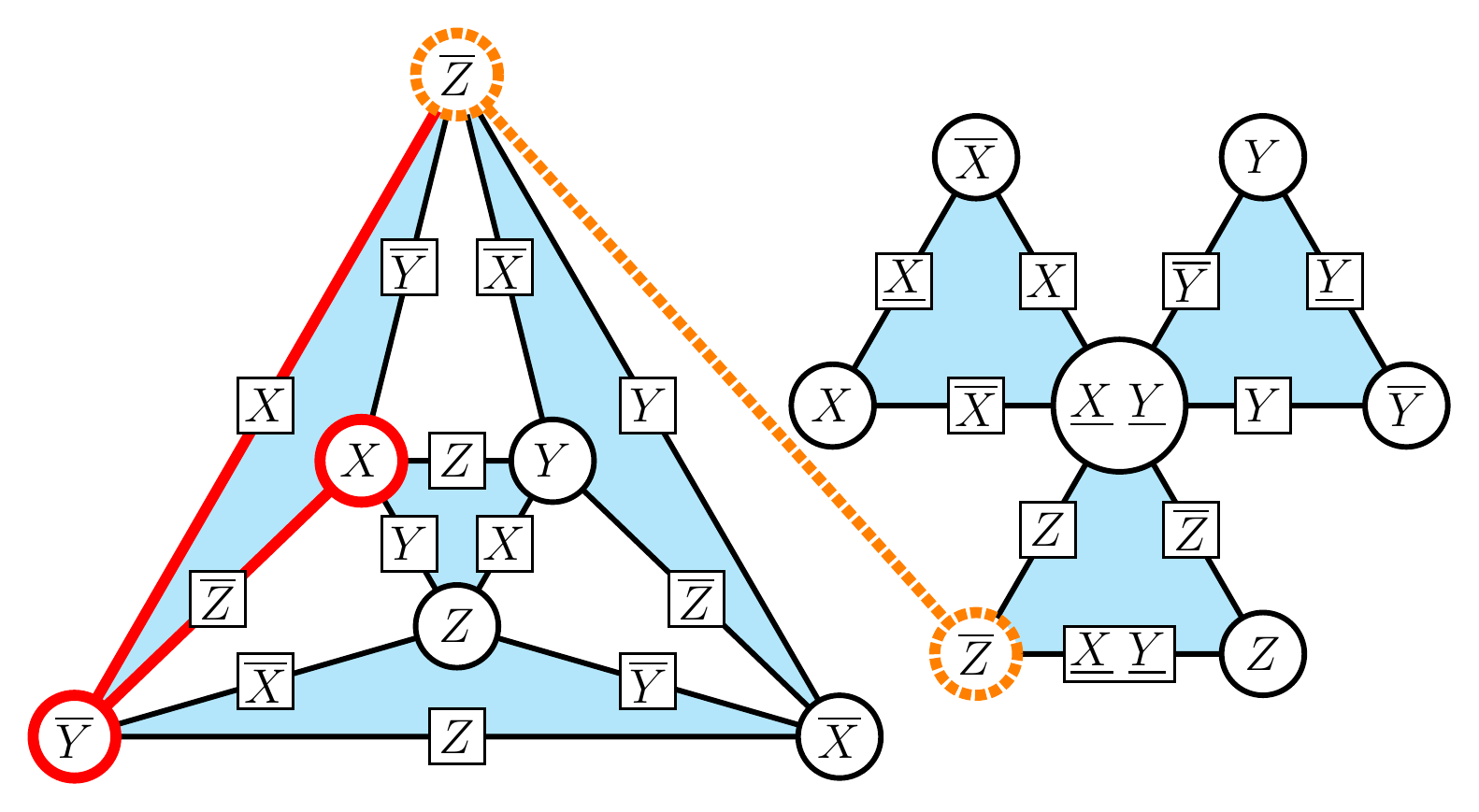}
\caption{A path attempting to find a doubling product operator that includes the operator string $X\zo X \zo \yu \yu$. In this case, the process fails when passing an edge labeled $X$, arriving at a vertex labeled $\zo$, and then attempting to find another vertex labeled with $\zo$. Thus, we conclude that $X\zo X\zo \yu \yu$ is not part of any doubling product operator.}
  \label{fig:comm_structure_pathfind_fail} 
\end{figure}

There are cases where no doubling product operator contains a given operator string. For example, consider $\bm{A}=X\zo X\zo \yu \yu$. Attempting to find a path that forms a doubling product operator containing $\bm{A}$ fails when reaching a vertex labeled $\zo$: since no edge from this vertex is labeled with $\zo$, it becomes impossible to complete the sequence. Fig.\ref{fig:comm_structure_pathfind_fail} illustrates this failure diagrammatically.

For any operator string $\bm{A}$ of length $k$, a path with $1\leq v\leq k-1$ vertices can be constructed. If the number of vertices $v$ in the path satisfies $1\leq v \leq k-2$, then the operator string $\bm{A}$ cannot be part of a doubling product operator When $v=k-1$, $\bm{A}$ is included in a doubling product operator if and only if the last operator $A_{k}$ of $\bm{A}$ appears in the final vertex of the path. 

We refer to a path that attempts to yield a doubling product operator containing $\bm{A}$ as a \textit{DP path} of $\bm{A}$, where DP stands for doubling product. Note that a unique DP path exists for any operator string $\bm{A}$, as established by previous arguments. If a DP path successfully forms a doubling product operator containing $\bm{A}$, we refer to it as a \textit{perfect DP path}.

\begin{figure}[t] 
\centering
\includegraphics[width=1\linewidth]{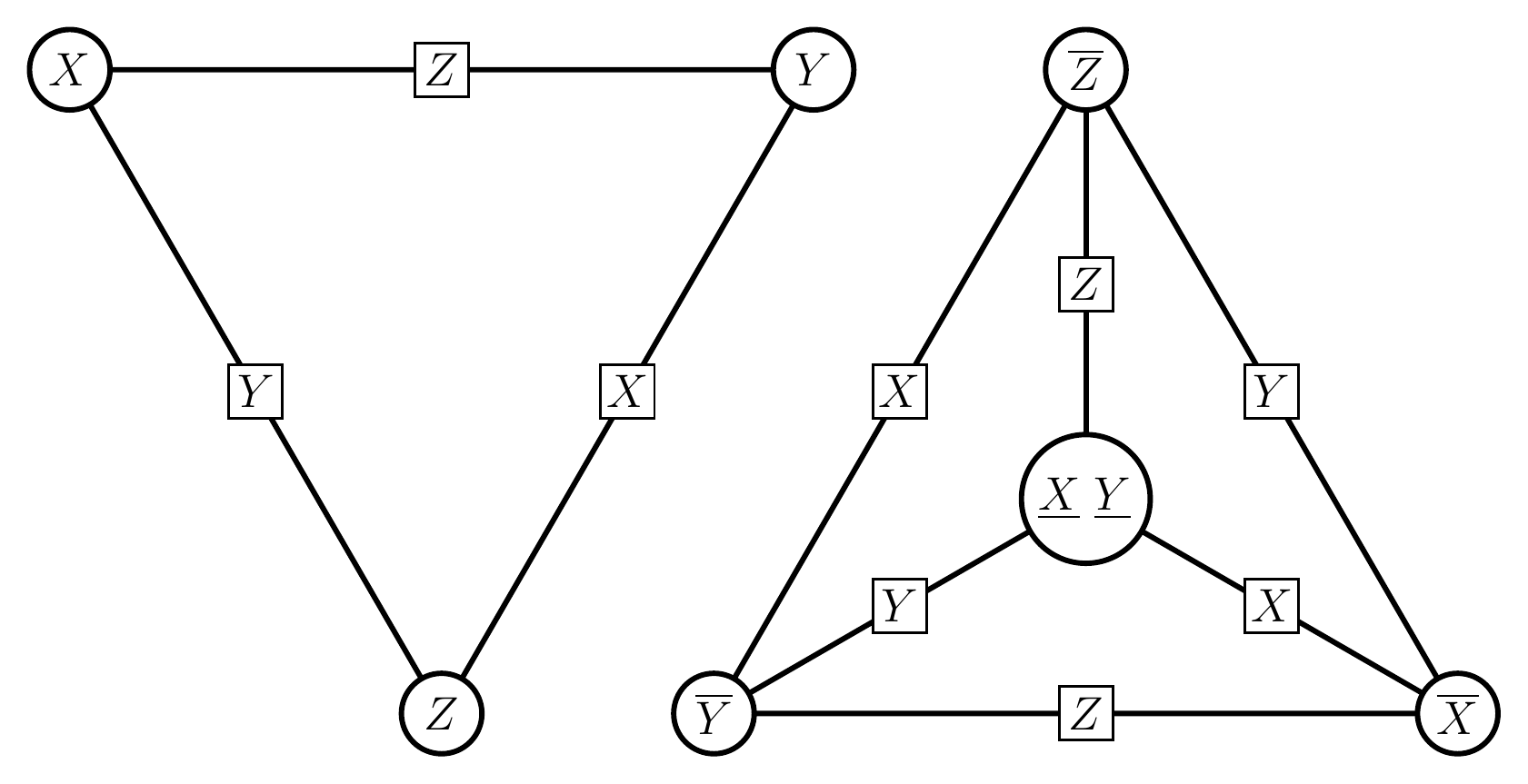}
\caption{Diagrams useful for proving Step 1 for the Hamiltonian $H_\theta$ in cases where certain coefficients are zero.}
  \label{fig:comm_structure_xyzdep} 
\end{figure}

One final note on this diagrammatic representation is that Fig.\ref{fig:comm_structure} is applicable only if the Hamiltonian
\begin{equation}
H_\theta=\sum_{1\leq j\leq L, A\in \mathcal{B}} J(A) A_j A_{j+1}
\end{equation}
satisfies $J(A)\neq 0$ for all $A\in \mathcal{B}$. If any of these coefficients are zero, a different diagram is required. For example, if only $J(X), J(Y),$ and $J(Z)$ are nonzero, then the left panel of Fig.\ref{fig:comm_structure_xyzdep} should be used. Conversely, if all coefficients are nonzero except $J(X)=J(Y)=J(Z)=0$, then the right panel of Fig.\ref{fig:comm_structure_xyzdep} applies. These diagrams are particularly useful when considering the Hamiltonian $H_\theta$ with $\theta=0$ or $\theta=\arctan 2$.

\section{Step 1: Detailed Proof}\label{section:3}
In this section, we present a detailed proof of the statement in Step 1 of our main text, which asserts that if $\len [C,H_{\theta}]\leq \len C$, then $C|_{\len C}$ must be a sum of doubling product operators with appropriate coefficients, $\gamma_\theta$, up to a scaling factor. The appropriate coefficient for a doubling product operator $\bm{D}^k = \ob{A_{(1)}\cdots A_{(k-1)}}$ is defined as
\begin{align*}
\gamma_\theta(\bm{D}) = \sum_{\ob{\scriptstyle B_{(1)}\cdots B_{(k-1)}}= \bm{D}} \prod_{n=1}^{k-2} \coeff ([B_{(n)}B_{(n+1)}])\cdot \prod_{n=1}^{k-1} J_\theta (B_{(n)})\numberthis \label{eq:appropriate_coeff}
\end{align*}
where
\begin{align*}
J_\theta (A)=\begin{cases}
J_1\coloneqq \cos\theta-\frac{1}{2}\sin\theta&A=X, Y, Z\\
J_2\coloneqq \frac{1}{3}\sin\theta&A=\xu,\yu,\zu\\
J_3\coloneqq \frac{1}{2}\sin\theta&A=\xo, \yo,\zo
\end{cases}
\end{align*}

As previously discussed, the set of doubling product operators varies depending on whether the coefficients $J_1, J_2,$ and $J_3$ are zero or non-zero. Therefore, we separate these cases and examine each one individually.

\subsection{Case 1: $J_1, J_2, J_3\neq 0$}
We begin by examining the case where all coefficients $J_1, J_2,$ and $J_3$ are nonzero, corresponding to $\theta\neq 0, \arctan 2$. The main ideas of the proof in this section remain valid for the cases $\theta=0$ or $\theta=\arctan 2$.

In the lemmas below, $\bm{A}$ denotes a length $k$ operator string within a length $k$ operator $C$, satisfying $\len([C,H])\leq k$. First, we demonstrate that if $\bm{A}$ has an imperfect DP path, then it must have zero coefficient.

\begin{lemma}\label{lemma:1}
Suppose that $\bm{A}$ has a DP path with only one vertex. Then $q(\bm{A})$ must be zero.
\end{lemma}
\begin{proof}
The assumption that $\bm{A}$ has a DP path with a single vertex implies that the first two operators of $\bm{A}$ must form one of the operator strings in the Hamiltonian $H_\theta$; that is, $A_{(1)}A_{(2)}\in \{XX, YY, ZZ, \xo\xo, \yo\yo, \zo\zo, \xu\xu, \xu\yu, \yu\xu, \yu\yu\}$. Now, assume $A_{(k)}\not\in \{\xu, \yu\}$. In this case, the following commutator holds:
\begin{equation*}
\begin{array}{ccccccccc}
 &A_{(1)}&A_{(2)}&A_{(3)}&\cdots&A_{(k-1)}&A_{(k)}     & \\
 &       &       &       &      &&U_{R}      &U_{R} \\\hline
 &A_{(1)}&A_{(2)}&A_{(3)}&\cdots&A_{(k-1)}& A_{(k)}/U_{R} &U_{R}
\end{array}
\end{equation*}
Here, $U_R=X$ if $A_{(k)}\in \{Y,Z,\xo,\yo,\zo\}$ and $U_R=Y$ when $A_{(k)}=X$, using the notation $/$ as defined in the main text. This indicates there is exactly one commutator that yields $A_{(1)}\cdots A_{(k-1)}A_{(k)}/U_R U_R$, a length $k+1$ operator string. Given the assumption $\len ([C,H])\leq k$, we conclude $p(A_{(1)}\cdots A_{(k)}/U_R U_R)=0$, which implies $q(A_{(1)}\cdots A_{(k)})=0$.

Now, consider the case $A_{(k)}=\xu$. In this case, we analyze the following commutators:
\begin{equation*}
\begin{array}{ccccccccc}
 &A_{(1)}&A_{(2)}&A_{(3)}&\cdots&A_{(k-1)}&\xu     & \\
 &       &       &       &      &&X      &X \\\hline
  &A_{(1)}&A_{(2)}&A_{(3)}&\cdots&A_{(k-1)}&\yu     & \\
 &       &       &       &      &&X      &X \\\hline
 &A_{(1)}&A_{(2)}&A_{(3)}&\cdots&A_{(k-1)}& \xo &X
\end{array}
\end{equation*}
This yields $2q(\bm{A}) - q(A_{(1)}\cdots A_{(k-1)}\yu)=0$. Now, considering the following commutators:
\begin{equation*}
\begin{array}{ccccccccc}
 &A_{(1)}&A_{(2)}&A_{(3)}&\cdots&A_{(k-1)}&\xu     & \\
 &       &       &       &      &&Y      &Y \\\hline
  &A_{(1)}&A_{(2)}&A_{(3)}&\cdots&A_{(k-1)}&\yu     & \\
 &       &       &       &      &&Y      &Y \\\hline
 &A_{(1)}&A_{(2)}&A_{(3)}&\cdots&A_{(k-1)}& \yo &Y
\end{array}
\end{equation*}
we find $2q(\bm{A}) - 2q(A_{(1)}\cdots A_{(k-1)}\yu)=0$. Combining these two equations gives $q(\bm{A}) = q(A_{(1)}\cdots A_{(k-1)}\yu) = 0$. Therefore, $q(\bm{A})=0$ if $A_{(k)}=\xu$; by similar reasoning, $q(\bm{A})=0$ if $A_{(k)}=\yu$. This completes the proof.
\end{proof}

\begin{lemma}\label{lemma:2}
Suppose that $\bm{A}$ has a DP path with $2\leq v\leq k-2$ vertices. Then $q(\bm{A})$ must be zero.
\end{lemma}
\begin{proof}
First, consider the case $A_{(k)}\not\in \{\xu,\yu\}$, and analyze the following commutators:
\begin{equation*}
\begin{array}{ccccccccc}
 &A_{(1)}&A_{(2)}&A_{(3)}&\cdots&A_{(k-1)}&A_{(k)}     & \\
 &       &       &       &      &&U_{R}      &U_{R} \\\hline
  &       &      U_{L}&A_{(3)}&\cdots&A_{(k-1)}& A_{(k)}/U_{R} &U_{R}\\
 &P_{L}&      P_{L}&       &      &&           & \\\hline
 &       &      U_{L}&A_{(3)}&\cdots&A_{(k-1)}& A_{(k)}/U_{R} &U_{R}\\
 &P_{L}&      P'_{L}&       &      &&           & \\\hline
  &       &      U'_{L}&A_{(3)}&\cdots&A_{(k-1)}& A_{(k)}/U_{R} &U_{R}\\
 &P_{L}&      P_{L}&       &      &&           & \\\hline
 &A_{(1)}&A_{(2)}&A_{(3)}&\cdots&A_{(k-1)}& A_{(k)}/U_{R} &U_{R}
\end{array}
\end{equation*}
Here, the primed notation indicates $\xu'=\yu$, $\yu'=\xu$, and $0$ for all the other operators. Again, choose $U_R=X$ when $A_{(k)}\in \{Y,Z,\xo,\yo,\zo\}$ and $U_R=Y$ when $A_{(k)}=X$. These commutators implies the following equation:
\begin{align*}
0 & = q(A_{(1)}\cdots A_{(k)}) \coeff([A_{(k)},U_R]) J_\theta(U_R) \\
&+ J_\theta(P_L P_L) \coeff([U_L,P_L]|A_{(2)}) q(U_L A_{(3)}\cdots A_{(k-1)} (A_{(k)}/U_R) U_R) \\
&+ J_\theta(P_L P'_L) \coeff([U_L,P'_L]|A_{(2)}) q(U_L A_{(3)}\cdots A_{(k-1)} (A_{(k)}/U_R) U_R) \\
&+ J_\theta(P_L P_L) \coeff([U'_L,P_L]|A_{(2)}) q(U'_L A_{(3)}\cdots A_{(k-1)} (A_{(k)}/U_R) U_R).
\end{align*}
Here, $J_\theta(P_LP_L)$ or $J_\theta(P_LP'_L)$ means the coefficient of those operator string in the Hamiltonian $H_\theta$ in \ref{eq:Ham_modifiedform}, and $\coeff([U_L,P_L]|A_{(2)})$ represents the contribution of $A_{(2)}$ to the commutator $[U_L,P_L]$; for example, $\coeff([Z,\zo]|\yu) = 2i$. Crucially, since $v\leq k-2$, the DP path of each operator string in this equation, except $\bm{A}$, has at most $v-1$ vertices, and by the induction hypothesis, their coefficients must be zero. Therefore, we conclude that $q(A_{(1)}\cdots A_{(k)})=0$.

The case $A_{(k)}\in \{\xu,\yu\}$ is very similar. Suppose $A_{(k)}=\xu$, and consider the following commutators:
\begin{equation*}
\begin{array}{ccccccccc}
 &A_{(1)}&A_{(2)}&A_{(3)}&\cdots&A_{(k-1)}&\xu     & \\
 &       &       &       &      &&X     &X \\\hline
  &A_{(1)}&A_{(2)}&A_{(3)}&\cdots&A_{(k-1)}&\yu     & \\
 &       &       &       &      &&X     &X \\\hline
  &       &      U_{L}&A_{(3)}&\cdots&A_{(k-1)}& \xo &X\\
 &P_{L}&      P_{L}&       &      &&           & \\\hline
 &       &      U_{L}&A_{(3)}&\cdots&A_{(k-1)}& \xo &X\\
 &P_{L}&      P'_{L}&       &      &&           & \\\hline
  &       &      U'_{L}&A_{(3)}&\cdots&A_{(k-1)}& \xo &X\\
 &P_{L}&      P_{L}&       &      &&           & \\\hline
 &A_{(1)}&A_{(2)}&A_{(3)}&\cdots&A_{(k-1)}& \xo &X
\end{array}
\end{equation*}
Every operator string in these commutators, except the first two, has a zero coefficient due to the induction hypothesis, giving $2q(\bm{A})  - q(A_{(1)}\cdots A_{(k-1)}\yu)=0$. Similarly, from the following commutators:
\begin{equation*}
\begin{array}{ccccccccc}
 &A_{(1)}&A_{(2)}&A_{(3)}&\cdots&A_{(k-1)}&\xu     & \\
 &       &       &       &      &&Y     &Y \\\hline
  &A_{(1)}&A_{(2)}&A_{(3)}&\cdots&A_{(k-1)}&\yu     & \\
 &       &       &       &      &&Y     &Y \\\hline
  &       &      U_{L}&A_{(3)}&\cdots&A_{(k-1)}& \yo &Y\\
 &P_{L}&      P_{L}&       &      &&           & \\\hline
 &       &      U_{L}&A_{(3)}&\cdots&A_{(k-1)}& \yo &Y\\
 &P_{L}&      P'_{L}&       &      &&           & \\\hline
  &       &      U'_{L}&A_{(3)}&\cdots&A_{(k-1)}& \yo &Y\\
 &P_{L}&      P_{L}&       &      &&           & \\\hline
 &A_{(1)}&A_{(2)}&A_{(3)}&\cdots&A_{(k-1)}& \yo &Y
\end{array}
\end{equation*}
we obtain $q(\bm{A})  - 2q(A_{(1)}\cdots A_{(k-1)}\yu)=0$. Solving these equations gives $q(\bm{A}) = q(A_{(1)}\cdots A_{(k-1)}\yu)=0$. A similar argument holds for $A_{(k)}=\yu$.
\end{proof}

\begin{lemma}\label{lemma:3}
Suppose that $\bm{A}$ has a DP path with $k-1$ vertices, but the operator in the last vertex of the DP path differs from $A_{(k)}$. Then $q(\bm{A})$ must be zero.
\end{lemma}
\begin{proof}
The assumption implies that the operator string $\bm{A}$ satisfies
\begin{equation*}
\bm{A}\leq \ob{D_{(1)}\cdots D_{(k-1)}}/U_R.
\end{equation*}
Here, $D_{(n)}$ is the operator in the $n$-th vertex of the DP path, and $U_R$ is an operator such that $A_{(k)}\leq [D_{(k-1)},U_R]$. Such a $U_R$ always exists since $A_{(k)}$ and $D_{(k-1)}$ are in different vertices in Fig.\ref{fig:comm_structure}.

Consider the case $A_{(k)}\not\in \{\xu,\yu\}$ and examine the following commutators:
\begin{equation*}
\begin{array}{ccccccccc}
 &A_{(1)}&A_{(2)}&A_{(3)}&\cdots&A_{(k-1)}&A_{(k)}     & \\
 &       &       &       &      &&U_{R}      &U_{R} \\\hline
  &A_{(1)}&A_{(2)}&A_{(3)}&\cdots&A_{(k-1)}&A_{(k)}     & \\
 &       &       &       &      &&U'_{R}      &U_{R} \\\hline
  &       &      U_{L}&A_{(3)}&\cdots&A_{(k-1)}& D_{(k-1)} &U_{R}\\
 &P_{L}&      P_{L}&       &      &&           & \\\hline
 &       &      U_{L}&A_{(3)}&\cdots&A_{(k-1)}& D_{(k-1)} &U_{R}\\
 &P_{L}&      P'_{L}&       &      &&           & \\\hline
  &       &      U'_{L}&A_{(3)}&\cdots&A_{(k-1)}& D_{(k-1)} &U_{R}\\
 &P_{L}&      P_{L}&       &      &&           & \\\hline
 &A_{(1)}&A_{(2)}&A_{(3)}&\cdots&A_{(k-1)}& D_{(k-1)} &U_{R}
\end{array}
\end{equation*}
Here, we used the fact that $A_{(k)}\leq [D_{(k-1)},U_R]$ implies $D_{(k-1)}\leq [A_{(k)},U_R]$, which follows from the commutator table or Fig.\ref{fig:comm_structure}. It is straightforward to verify that the DP path for every operator string in these commutators, except $\bm{A}$, has at most $k-2$ vertices. By Lemma \ref{lemma:2}, these operator strings have zero coefficients.

If $U_R\not\in \{\xu,\yu\}$, the second commutator is trivial, leading to $q(\bm{A})=0$. If $U_R\in \{\xu,\yu\}$, then both commutators yield nontrivial contributions resulting in the equation
\begin{align*}
&q(\bm{A})\cdot\\
&\cdot(\coeff([A_{(k)},U_R]|D_{(k-1)})J_\theta(U_RU_R)+\coeff([A_{(k)},U'_R]|D_{(k-1)})J_\theta(U'_RU_R))=0
\end{align*}
Using $J_\theta(U_RU_R)=2J_2$ and $J_\theta(U_R'U_R)=J_2$, we obtain
\begin{equation*}
J_2 q(\bm{A})(2\coeff([A_{(k)},U_R]|D_{(k-1)})+\coeff([A_{(k)},U'_R]|D_{(k-1)}))=0.
\end{equation*}
If $U_R\in \{\xu,\yu\}$, we can choose $U_R$ as either $\xu$ or $\yu$: if $A_{(k)}\in \{X,\xo, Z,\zo\}$, select $U_R=\xu$; if $A_{(k)}\in \{Y,\yo \}$, select $U_R=\yu$. This choice ensures that 
\begin{equation*}
2\coeff([A_{(k)},U_R]|D_{(k-1)})+\coeff([A_{(k)},U'_R]|D_{(k-1)}\neq 0,
\end{equation*}
which implies $q(\bm{A})=0$.

Finally, consider the case $A_{(k)}\in \{\xu,\yu\}$. Let $A_{(k)}=\xu$, and analyze the following commutators:
\begin{equation*}
\begin{array}{ccccccccc}
 &A_{(1)}&A_{(2)}&A_{(3)}&\cdots&A_{(k-1)}&\xu     & \\
 &       &       &       &      &&X      &X \\\hline
  &A_{(1)}&A_{(2)}&A_{(3)}&\cdots&A_{(k-1)}&\yu     & \\
 &       &       &       &      &&X      &X \\\hline
  &       &      U_{L}&A_{(3)}&\cdots&A_{(k-1)}& \xo &X \\
 &P_{L}&      P_{L}&       &      &&           & \\\hline
 &       &      U_{L}&A_{(3)}&\cdots&A_{(k-1)}& \xo &X \\
 &P_{L}&      P'_{L}&       &      &&           & \\\hline
  &       &      U'_{L}&A_{(3)}&\cdots&A_{(k-1)}& \xo &X \\
 &P_{L}&      P_{L}&       &      &&           & \\\hline
 &A_{(1)}&A_{(2)}&A_{(3)}&\cdots&A_{(k-1)}& \xo &X 
\end{array}
\end{equation*}
Here, every operator string, except $\bm{A}$ and $A_{(1)}\cdots A_{(k-1)}\yu$, does not end with $\xu$ or $\yu$ and has an imperfect DP path. The argument above shows that these operator strings have zero coefficients, giving $2q(\bm{A}) - q(A_{(1)}\cdots A_{(k-1)}\yu)=0$. Similarly, the following commutators:
\begin{equation*}
\begin{array}{ccccccccc}
 &A_{(1)}&A_{(2)}&A_{(3)}&\cdots&A_{(k-1)}&\xu     & \\
 &       &       &       &      &&Y      &Y \\\hline
  &A_{(1)}&A_{(2)}&A_{(3)}&\cdots&A_{(k-1)}&\yu     & \\
 &       &       &       &      &&Y      &Y \\\hline
  &       &      U_{L}&A_{(3)}&\cdots&A_{(k-1)}& \yo &Y \\
 &P_{L}&      P_{L}&       &      &&           & \\\hline
 &       &      U_{L}&A_{(3)}&\cdots&A_{(k-1)}& \yo &Y \\
 &P_{L}&      P'_{L}&       &      &&           & \\\hline
  &       &      U'_{L}&A_{(3)}&\cdots&A_{(k-1)}& \yo &Y \\
 &P_{L}&      P_{L}&       &      &&           & \\\hline
 &A_{(1)}&A_{(2)}&A_{(3)}&\cdots&A_{(k-1)}& \yo &Y 
\end{array}
\end{equation*}
give $q(\bm{A}) - 2q(A_{(1)}\cdots A_{(k-1)}\yu)=0$. Combining these equations, we find $q(\bm{A}) = q(A_{(1)}\cdots A_{(k-1)}\yu)=0$. The case $A_{(k)}=\yu$ can be treated similarly.
\end{proof}

Lemmas \ref{lemma:1}, \ref{lemma:2}, and \ref{lemma:3} establish that if $\bm{A}$ does not have a perfect DP path, i.e., if $\bm{A}$ is not part of a doubling product operator, then $q(\bm{A})=0$. Our next goal is to determine the coefficients of the operator strings contained within the doubling product operators.

Consider the operator string $\ob{(XY)^{m-1}X}$ for even $k=2m$ and $\ob{(XY)^{m}}$ for odd $k=2m+1$. We denote these strings as $\bm{A}_{\text{base}}$. Note that $\bm{A}_{\text{base}}$ can be expressed in the doubling product form.

\begin{lemma}\label{lemma:4}
Suppose an operator string $\bm{A}$ is part of a doubling product operator. Additionally, assume the coefficient of $\bm{A}_{\text{base}}$ in $C$ is fixed. Then the coefficient $q(\bm{A})$ of $\bm{A}$ in $C$ is uniquely determined.
\end{lemma}
\begin{proof}
Suppose $\bm{A}$ does not end with $\xu$ or $\yu$, and consider the following commutators:
\begin{equation*}
\begin{array}{ccccccccc}
 &A_{(1)}&A_{(2)}&A_{(3)}&\cdots&A_{(k-1)}&A_{(k)}    & \\
 &       &       &       &      &&A_{(k)}/P_1       &P_1 \\\hline
  &       &      U_{L}&A_{(3)}&\cdots&A_{(k-1)}&A_{(k)}/P_1       &P_1  \\
 &P_{L}&      P_{L}&       &      &&      & \\\hline
 &       &      U_{L}&A_{(3)}&\cdots&A_{(k-1)}&A_{(k)}/P_1        &P_1 \\
 &P_{L}&      P'_{L}&       &      &&       & \\\hline
  &       &      U'_{L}&A_{(3)}&\cdots&A_{(k-1)}&A_{(k)}/P_1          &P_1 \\
 &P_{L}&      P_{L}&       &      &&       & \\\hline
 &A_{(1)}&A_{(2)}&A_{(3)}&\cdots&A_{(k-1)}&A_{(k)}/P_1        &P_1 
\end{array}
\end{equation*}
Here $P_1=X$ if $A_{(k)}\in \{Y,Z,\xo,\yo,\zo\}$ and $P_2=Y$ if $A_{(k)}=X$.

If $U_L\not\in \{\xu, \yu\}$, the last commutator does not contribute. Additionally, both the second and third commutators contain the operator string $U_LA_{(3)}\cdots A_{(k-1)}A_{(k)}/\ob{P_1}$. This setup provides a linear relationship between $q(U_LA_{(3)}\cdots A_{(k-1)}A_{(k)}/\ob{P_1})$ and $q(\bm{A})$.

Now, consider the case where $U_L = \xu$. In this case, the third commutator does not contribute, and we need to consider the following three commutators:
\begin{equation*}
\begin{array}{ccccccccc}
 &A_{(1)}&A_{(2)}&A_{(3)}&\cdots&A_{(k-1)}&A_{(k)}    & \\
 &       &       &       &      &&[A_{(k)},P_1]      &P_1 \\\hline
  &       &      \xu&A_{(3)}&\cdots&A_{(k-1)}&A_{(k)}/P_1        &P_1  \\
 &P_{L}&      P_{L}&       &      &&      & \\\hline
 &       &      \yu&A_{(3)}&\cdots&A_{(k-1)}&A_{(k)}/P_1         &P_1 \\
 &P_{L}&      P_{L}&       &      &&       & \\\hline
 &A_{(1)}&A_{(2)}&A_{(3)}&\cdots&A_{(k-1)}&A_{(k)}/P_1         &P_1 
\end{array}
\end{equation*}
This setup yields a linear equation of the coefficients $q(\xu A_{(3)}\cdots A_{(k)}/\ob{P_1})$, $q(\yu A_{(3)}\cdots A_{(k)}/\ob{P_1})$, and $q(\bm{A})$. By applying the same argument as above, we establish a linear relation between $q(U_2A_{(4)}\cdots A_{(k)}/\ob{P_1}/\ob{P_2})$ and $q(\xu A_{(3)}\cdots A_{(k)}/\ob{P_1})$, as well as between $q(U_2A_{(4)}\cdots A_{(k)}/\ob{P_1}/\ob{P_2})$ and $q(\yu A_{(3)}\cdots A_{(k)}/\ob{P_1})$. Here $U_2$ satisfies $A_{(3)}\leq [U_2, \xu]$, and $P_2=Y$ when $P_1=X$, and $P_2=X$ when $P_1=Y$. By combining these linear relationships, we obtain a direct linear relation between $q(U_2A_{(4)}\cdots A_{(k)}/\ob{P_1}/\ob{P_2})$ and $q(\bm{A})$. Similar argue holds for $U_L=\yu$.

By repeatedly applying this argument, we continue to establish proportional relationships between the coefficients, such as
\begin{equation*}
q(\bm{A})\sim q(U_2 A_{(4)}\cdots A_{(k)}/\ob{P_1 P_2}) \sim q(U_3 A_{(5)}\cdots A_{(k)}/\ob{P_1 P_2 P_1}) \sim \cdots 
\end{equation*}
where $\sim$ indicates proportionality between two variables. Since $\bm{A}$ is part of a doubling product operator, this argument can be extended until we reach $\bm{A}_{\text{base}}$. Therefore,
\begin{equation}
q(\bm{A}) \sim q(\bm{A}_{\text{base}}),
\end{equation}
which implies that if $q(\bm{A}_{\text{base}})$ is fixed, then $q(\bm{A})$ is also fixed.

Now, suppose that $\bm{A}$ ends with $\xu$, and consider the following commutators.
\begin{equation*}
\begin{array}{ccccccccc}
 &A_{(1)}&A_{(2)}&A_{(3)}&\cdots&A_{(k-1)}&\xu   & \\
 &       &       &       &      &&X       &X \\\hline
  &A_{(1)}&A_{(2)}&A_{(3)}&\cdots&A_{(k-1)}&\yu   & \\
 &       &       &       &      &&X       &X \\\hline
  &       &      U_{L}&A_{(3)}&\cdots&A_{(k-1)}&\xo       &X  \\
 &P_{L}&      P_{L}&       &      &&      & \\\hline
 &       &      U_{L}&A_{(3)}&\cdots&A_{(k-1)}&\xo       &X \\
 &P_{L}&      P'_{L}&       &      &&       & \\\hline
  &       &      U'_{L}&A_{(3)}&\cdots&A_{(k-1)}&\xo       &X \\
 &P_{L}&      P_{L}&       &      &&       & \\\hline
 &A_{(1)}&A_{(2)}&A_{(3)}&\cdots&A_{(k-1)}&\xo       &X
\end{array}
\end{equation*}
We have already shown that the operator strings in the last three commutators are uniquely determined when $q(\bm{A}_{\text{base}})$ is fixed. Therefore, we obtain
\begin{equation*}
2q(\bm{A})-q(A_{(1)}\cdots \yu) = c_1 q(\bm{A}_{\text{base}})
\end{equation*}
for some constant $c_1$. Similarly, from the following commutators:
\begin{equation*}
\begin{array}{ccccccccc}
 &A_{(1)}&A_{(2)}&A_{(3)}&\cdots&A_{(k-1)}&\xu   & \\
 &       &       &       &      &&Y       &Y \\\hline
  &A_{(1)}&A_{(2)}&A_{(3)}&\cdots&A_{(k-1)}&\yu   & \\
 &       &       &       &      &&Y       &Y \\\hline
  &       &      U_{L}&A_{(3)}&\cdots&A_{(k-1)}&\yo       &Y  \\
 &P_{L}&      P_{L}&       &      &&      & \\\hline
 &       &      U_{L}&A_{(3)}&\cdots&A_{(k-1)}&\yo       &Y \\
 &P_{L}&      P'_{L}&       &      &&       & \\\hline
  &       &      U'_{L}&A_{(3)}&\cdots&A_{(k-1)}&\yo       &Y \\
 &P_{L}&      P_{L}&       &      &&       & \\\hline
 &A_{(1)}&A_{(2)}&A_{(3)}&\cdots&A_{(k-1)}&\yo       &Y
\end{array}
\end{equation*}
one get
\begin{equation*}
-q(\bm{A})+2q(A_{(1)}\cdots A_{(k-1)} \yu) = c_2 q(\bm{A}_{\text{base}})
\end{equation*}
for some constant $c_2$. Using these two equations, we find
\begin{align*}
q(\bm{A})&= \frac{2c_1+c_2}{3}q(\bm{A}_{\text{base}})\\
q(A_{(1)}\cdots A_{(k-1)} \yu)&= \frac{c_1+2c_2}{3}q(\bm{A}_{\text{base}})
\end{align*}
showing that $q(\bm{A})$ is uniquely determined when $q(\bm{A}_{\text{base}})$ is fixed. The same argument applies for an operator string $\bm{A}$ that ends with $\yu$. This completes the proof of the claim.
\end{proof}

To conclude this subsection, we identify the operator $C_{\text{boost}}$ of length $k$ that satisfies $\len([C_{\text{boost}},H_\theta])\leq k$. The length-$k$ component $C_{\text{boost}}|_k$ defines the coefficients in Eq.\ref{eq:appropriate_coeff}. Together with Lemma \ref{lemma:4}, this completes the proof of our statement in Step 1 of the main text.

\begin{lemma}\label{lemma:5}
Let the Hamiltonian be $H_\theta=\sum_{j}h_{j}$, where $h_{j}$ is the term acting on sites $j$ and $j+1$. Define
\begin{equation}
C_{\text{boost}}= \sum_j [h_{j},[h_{j+1},[\cdots ,[h_{j+k-4},[h_{j+k-3},h_{j+k-2}]]\cdots]]].
\end{equation}
Then $\len([C_{\text{boost}},H_\theta])\leq k$.
\end{lemma}
\begin{proof}
By calculating $[C_{\text{boost}},H_\theta]$ and isolating terms acting on $k+1$ consecutive sites, we find
\begin{align*}
[C_{\text{boost}},H_\theta]|_{k+1} &= \sum_j [[h_{j},[h_{j+1},[\cdots ,[h_{j+k-4},[h_{j+k-3},h_{j+k-2}]]\cdots]]],h_{j+k-1}]\\
&+\sum_j [[h_{j},[h_{j+1},[\cdots ,[h_{j+k-4},[h_{j+k-3},h_{j+k-2}]]\cdots]]],h_{j-1}]\\
&= \sum_j [[h_{j},[h_{j+1},[\cdots ,[h_{j+k-4},[h_{j+k-3},h_{j+k-2}]]\cdots]],h_{j+k-1}]]\\
&+\sum_j [[h_{j},[h_{j+1},[\cdots ,[h_{j+k-4},[h_{j+k-3},h_{j+k-2}]]\cdots]]],h_{j-1}]\\
&= \sum_j [[h_{j},[h_{j+1},[\cdots ,[h_{j+k-4},[h_{j+k-3},h_{j+k-2}]]\cdots],h_{j+k-1}]]]\\
&+\sum_j [[h_{j},[h_{j+1},[\cdots ,[h_{j+k-4},[h_{j+k-3},h_{j+k-2}]]\cdots]]],h_{j-1}]\\
&=\cdots\\
&= \sum_j [h_{j},[h_{j+1},[\cdots ,[h_{j+k-4},[h_{j+k-3},[h_{j+k-2},h_{j+k-1}]]]\cdots]]]\\
&+\sum_j [[h_{j},[h_{j+1},[\cdots ,[h_{j+k-4},[h_{j+k-3},h_{j+k-2}]]\cdots]]],h_{j-1}]\\
&= \sum_j [h_{j},[h_{j+1},[\cdots ,[h_{j+k-4},[h_{j+k-3},[h_{j+k-2},h_{j+k-1}]]]\cdots]]]\\
&-\sum_j [h_{j-1},[h_{j},[h_{j+1},[\cdots ,[h_{j+k-4},[h_{j+k-3},h_{j+k-2}]]\cdots]]]]\\
&= \sum_j [h_{j},[h_{j+1},[\cdots ,[h_{j+k-4},[h_{j+k-3},[h_{j+k-2},h_{j+k-1}]]]\cdots]]]\\
&-\sum_j [h_{j},[h_{j+1},[\cdots ,[h_{j+k-4},[h_{j+k-3},[h_{j+k-2},h_{j+k-1}]]]\cdots]]]\\
&=0.
\end{align*}
Here we used the fact that $h_{j+k-1}$ commutes with $h_j,\cdots,h_{j+k-3}$ and employed the translational invariance of $H_\theta$. This confirms that $\len([C_{\text{boost}},H_\theta])\leq k$.
\end{proof}

\begin{lemma}\label{lemma:6}
$C_{\text{boost}}$ is equal to $\sum_{\bm{D}, j} \gamma_\theta(\bm{D}) \{\bm{D}\}_j$.
\end{lemma}
\begin{proof}
The construction of $C_{\text{boost}}$ aligns directly with the construction of the doubling product operator. Specifically, we have
\begin{equation}
h_j = \sum_{B\in \bm{B}} J_\theta(B)\{BB\}_j
\end{equation}
where $h_j$ is defined in \ref{lemma:5}. This implies that
\begin{equation}
C_{\text{boost}} = \sum_j \prod_{n=1}^{k-2} \coeff([B_{(n)},B_{(n+1)}]) \cdot \prod_{n=1}^{k-1} J_\theta (B_{(n)}) \cdot \{\ob{B_{(1)}\cdots B_{(k-1)}}\}_j 
\end{equation}
By grouping different doubling product forms that yield the same doubling product operators, we obtain
\begin{equation}
C_{\text{boost}} = \sum_{\bm{D},j} \gamma_\theta(\bm{D})\cdot \{\bm{D}\}_j,
\end{equation}
which proves the desired result.
\end{proof}

Thus, we have shown that if $J_1, J_2, J_3\neq 0$, then $\len([C,H_\theta]) \leq \len C$ implies $C|_k = \alpha C_{\text{boost}}$.

\subsection{Case 2: $J_2, J_3=0$}
In the case where $J_2, J_3=0$ while $J_1\neq 0$, i.e. $\theta=0$, the proof proceeds in a similar manner. In this scenario, the DP path should be constructed using the left diagram in Fig.\ref{fig:comm_structure_xyzdep}. An important distinction is that, in this case, an operator string $\bm{A}$ may lack a DP path entirely if it does not begin with $X,Y,Z$.

As in the previous lemmas, $\bm{A}$ denotes a length $k$ operator string within a length $k$ operator $C$, satisfying $\len([C,H])\leq k$.

\begin{lemma}\label{lemma:7}
Suppose that $\bm{A}$ does not have a DP path or has a DP path with only one vertex. Then $q(\bm{A})$ must be zero.
\end{lemma}
\begin{proof}
Assume first that $A_{(k)}\not\in \{\xu, \yu\}$. In this case, we consider the following commutator:
\begin{equation*}
\begin{array}{ccccccccc}
 &A_{(1)}&A_{(2)}&A_{(3)}&\cdots&A_{(k-1)}&A_{(k)}     & \\
 &       &       &       &      &&U_{R}      &U_{R} \\\hline
 &A_{(1)}&A_{(2)}&A_{(3)}&\cdots&A_{(k-1)}& A_{(k)}/U_{R} &U_{R}
\end{array}
\end{equation*}
Here, $U_R=X$ when $A_{(k)}\in \{Y,Z,\xo,\yo,\zo\}$ and $U_R=Y$ when $A_{(k)}=X$. This is the only commutator that yields $A_{(1)}\cdots A_{(k-1)}A_{(k)}/U_{R}U_{R}$, so $q(\bm{A})=0$.

Now suppose $A_{(k)}= \xu$. In this case, consider the following commutators:
\begin{equation*}
\begin{array}{ccccccccc}
 &A_{(1)}&A_{(2)}&A_{(3)}&\cdots&A_{(k-1)}&\xu     & \\
 &       &       &       &      &&X      &X \\\hline
  &A_{(1)}&A_{(2)}&A_{(3)}&\cdots&A_{(k-1)}&\yu     & \\
 &       &       &       &      &&X      &X \\\hline
 &A_{(1)}&A_{(2)}&A_{(3)}&\cdots&A_{(k-1)}& \xo &X
\end{array}
\end{equation*}
and
\begin{equation*}
\begin{array}{ccccccccc}
 &A_{(1)}&A_{(2)}&A_{(3)}&\cdots&A_{(k-1)}&\xu     & \\
 &       &       &       &      &&Y      &Y \\\hline
  &A_{(1)}&A_{(2)}&A_{(3)}&\cdots&A_{(k-1)}&\yu     & \\
 &       &       &       &      &&Y      &Y \\\hline
 &A_{(1)}&A_{(2)}&A_{(3)}&\cdots&A_{(k-1)}& \yo &Y
\end{array}
\end{equation*}
These relations give $2q(\bm{A}) - q(A_{(1)}\cdots q_{(k-1)}\yu) = 0$ and $q(\bm{A}) - 2q(A_{(1)}\cdots q_{(k-1)}\yu) = 0$. Solving these equations give $q(\bm{A})=0$. Similarly, if $A_{(k)}=\yu$, then $q(\bm{A})=0$.
\end{proof}

\begin{lemma}
Suppose that $\bm{A}$ has a DP path with $2\leq v\leq k-2$ vertices. Then $q(\bm{A})$ must be zero.
\end{lemma}
\begin{proof}
By Lemma \ref{lemma:7} and the inversion symmetry of the Hamiltonian $H_0$, we can assume that $\bm{A}$ always starts and ends with $X,Y,$ or $Z$. Consider the following commutators:
\begin{equation*}
\begin{array}{ccccccccc}
 &A_{(1)}&A_{(2)}&A_{(3)}&\cdots&A_{(k-1)}&A_{(k)}     & \\
 &       &       &       &      &&U_{R}      &U_{R} \\\hline
  &       &      U_{L}&A_{(3)}&\cdots&A_{(k-1)}& A_{(k)}/U_{R} &U_{R}\\
 &P_{L}&      P_{L}&       &      &&           & \\\hline
  &       &      U'_{L}&A_{(3)}&\cdots&A_{(k-1)}& A_{(k)}/U_{R} &U_{R}\\
 &P_{L}&      P_{L}&       &      &&           & \\\hline
 &A_{(1)}&A_{(2)}&A_{(3)}&\cdots&A_{(k-1)}& A_{(k)}/U_{R} &U_{R}
\end{array}
\end{equation*}
Here, $P_{L}\in \{X,Y,Z\}$. If the third contribution is nontrivial, then $U_L\in \{\xu,\yu\}$, which causes the coefficients of the operator strings in the second and third commutators to be zero. Thus, it is sufficient to consider only the second contribution as nontrivial, establishing a linear relationship between $q(U_LA_{(3)}\cdots A_{(k-1)} A_{(k)}/U_R U_R)$ and $q(\bm{A})$. 

Since the DP path of $\bm{A}$ has $v$ vertices, the DP path of $U_LA_{(3)}\cdots A_{(k-1)} A_{(k)}/U_R U_R$ has $v-1$ vertices. By the induction hypothesis, $q(U_LA_{(3)}\cdots A_{(k-1)} A_{(k)}/U_R U_R)=0$, and therefore $q(\bm{A})=0$.
\end{proof}

\begin{lemma}
Suppose that $\bm{A}$ has a DP path with $k-1$ vertices, but the operator in the last vertex of the DP path is differs from $A_{(k)}$. Then $q(\bm{A})$ must be zero.
\end{lemma}
\begin{proof}
This assumption implies that the operator string $\bm{A}$ satisfies
\begin{equation}
\bm{A}\leq \ob{D_{(1)}\cdots D_{(k-1)}}/U_R.
\end{equation}
Here, $D_{(n)}$ is an operator in the $n$-th vertex of the DP path, and $U_R$ is an operator such that $A_{(k)}\leq [D_{(k-1)},U_R]$. We can assume $A_{(k)}\in \{X,Y,Z\}$, and since $A_{(k)}$ and $D_{(k-1)}$ are different, we can always find such a $U_R$ from $\{X,Y,Z\}$. Now, consider the following commutators:
\begin{equation*}
\begin{array}{ccccccccc}
 &A_{(1)}&A_{(2)}&A_{(3)}&\cdots&A_{(k-1)}&A_{(k)}     & \\
 &       &       &       &      &&U_{R}      &U_{R} \\\hline
  &       &      D_{(2)}&A_{(3)}&\cdots&A_{(k-1)}& D_{(k-1)} &U_{R}\\
 &D_{(1)}&      D_{(1)}&       &      &&           & \\\hline
 &A_{(1)}&A_{(2)}&A_{(3)}&\cdots&A_{(k-1)}& D_{(k-1)} &U_{R}
\end{array}
\end{equation*}
This construction establishes a linear relationship between $q(\bm{A})$ and $q(U_L A_{(3)}\cdots A_{(k-1)} D_{(k-1)}U_R)$. Due to the construction, the operator string $U_L A_{(3)}\cdots A_{(k-1)} D_{(k-1)}U_R$ has a DP path with $k-2$ vertices, which we have already shown to have a zero coefficient. Therefore, $q(\bm{A})=0$.
\end{proof}

Now consider the operator string $\ob{(XY)^{m-1}X}$ for even $k=2m$ and $\ob{(XY)^{m}}$ for odd $k=2m+1$, which are the same operator strings examined in the case where $J_1,J_2,J_3\neq 0$. We again denote them as $\bm{A}_{\text{base}}$.

\begin{lemma}
Suppose that an operator string $\bm{A}$ is contained in a doubling product operator. Additionally, suppose that the coefficient of $\bm{A}_{\text{base}}$ in $C$ is fixed. Then the coefficient $q(\bm{A})$ of $\bm{A}$ in $C$ is uniquely determined.
\end{lemma}
\begin{proof}
Note that in this case, $\bm{A}=\ob{D_{(1)}\cdots D_{(k-1)}}$. Consider the following commutators:
\begin{equation*}
\begin{array}{ccccccccc}
 &A_{(1)}&A_{(2)}&A_{(3)}&\cdots&A_{(k-1)}&A_{(k)}    & \\
 &       &       &       &      &&A_{(k)}/P_1       &P_1 \\\hline
  &       &      D_{(2)}&A_{(3)}&\cdots&A_{(k-1)}&A_{(k)}/P_1       &P_1  \\
 &D_{(1)}&      D_{(1)}&       &      &&      & \\\hline
 &A_{(1)}&A_{(2)}&A_{(3)}&\cdots&A_{(k-1)}&A_{(k)}/P_1        &P_1 
\end{array}
\end{equation*}
Here $P_1=X$ when $A_{(k)}\in \{Y,Z\}$ and $P_1=Y$ when $A_{(k)}=X$. These commutators show the linear relationship between $q(\bm{A})$ and $q(\ob{D_{(2)}\cdots D_{(k-1)} P_1})$. In a similar way, we can establish a linear relation between $q(\ob{D_{(2)}\cdots D_{(k-1)}P_1})$ and $q(\ob{D_{(3)}\cdots D_{(k-1)}P_1P_2})$, and so on. By iteratively finding these linear relationships, we eventually obtain a relationship between $q(\bm{A})$ and $q_{\bm{A}_{\text{base}}}$, showing that if $q_{\bm{A}_{\text{base}}}$ is fixed, then $q(\bm{A})$ is uniquely determined.
\end{proof}

Note that Lemmas \ref{lemma:5} and Lemma \ref{lemma:6} hold regardless of the values of $J_1, J_2,$ and $J_3$. Therefore, we have fully demonstrated that if $J_2=J_3=0$ and $J_1\neq 0$, i.e., $\theta=0$, then $\len ([C,H_0])\leq \len C$ implies that $C|_k= \alpha C_{\text{boost}}$. 

\subsection{Case 3: $J_1=0$}
When $J_1=0$ and $J_2,J_3\neq 0$, i.e., $\theta = \arctan 2$, the same logic we used before applies, with some minor modifications. In this case, we consider the DP path in the right diagram of Fig.\ref{fig:comm_structure_xyzdep}, noting that an operator string $\bm{A}$ may lack a DP path if it starts with $X,Y,$ or $Z$.

\begin{lemma}
Suppose that $\bm{A}$ has no DP path or has a DP path with only one vertex. Then $q(\bm{A})$ must be zero.
\end{lemma}
\begin{proof}
Assume $A_{(k)}\not\in \{\xu,\yu\}$. In this case consider the following commutator:
\begin{equation*}
\begin{array}{ccccccccc}
 &A_{(1)}&A_{(2)}&A_{(3)}&\cdots&A_{(k-1)}&A_{(k)}     & \\
 &       &       &       &      &&U_{R}      &U_{R} \\\hline
 &A_{(1)}&A_{(2)}&A_{(3)}&\cdots&A_{(k-1)}& A_{(k)}/U_{R} &U_{R}
\end{array}
\end{equation*}
Here $U_R=\xo$ when $A_{(k)}\in \{\yo,\zo\}$ and $U_R=\yo$ when $A_{(k)}=\xo$. We used the slash $/$ notation as defined in the main text. This unique commutator implies that $q(A_{(1)}\cdots A_{(k)})=0$.

Now suppose that $A_{(k)}=\xu$. In this case, consider the following commutators:
\begin{equation*}
\begin{array}{ccccccccc}
 &A_{(1)}&A_{(2)}&A_{(3)}&\cdots&A_{(k-1)}&\xu     & \\
 &       &       &       &      &&\xo      &\xo \\\hline
  &A_{(1)}&A_{(2)}&A_{(3)}&\cdots&A_{(k-1)}&\yu     & \\
 &       &       &       &      &&\xo      &\xo \\\hline
 &A_{(1)}&A_{(2)}&A_{(3)}&\cdots&A_{(k-1)}& X &\xo
\end{array}
\end{equation*}
This gives the relation $2q(\bm{A}) - q(A_{(1)}\cdots A_{(k-1)}\yu)=0$. Next, consider the following commutators:
\begin{equation*}
\begin{array}{ccccccccc}
 &A_{(1)}&A_{(2)}&A_{(3)}&\cdots&A_{(k-1)}&\xu     & \\
 &       &       &       &      &&\yo     &\yo \\\hline
  &A_{(1)}&A_{(2)}&A_{(3)}&\cdots&A_{(k-1)}&\yu     & \\
 &       &       &       &      &&\yo      &\yo \\\hline
 &A_{(1)}&A_{(2)}&A_{(3)}&\cdots&A_{(k-1)}& Y &\yo
\end{array}
\end{equation*}
This gives the relation $2q(\bm{A}) - 2q(A_{(1)}\cdots A_{(k-1)}\yu)=0$. Combining these two equations yields $q(\bm{A}) = q(A_{(1)}\cdots A_{(k-1)}\yu) = 0$. Therefore, $q(\bm{A})=0$ if $A_{(k)}=\xu$; by a similar argument, $q(\bm{A})=0$ if $A_{(k)}=\yu$. This completes the proof.
\end{proof}

\begin{lemma}\label{lemma:12}
Suppose that $\bm{A}$ has a DP path with $2\leq v\leq k-2$ vertices. Then $q(\bm{A})$ must be zero.
\end{lemma}
\begin{proof}
First, consider the case $A_{(k)}\not\in \{\xu,\yu\}$ and examine the following commutators:
\begin{equation*}
\begin{array}{ccccccccc}
 &A_{(1)}&A_{(2)}&A_{(3)}&\cdots&A_{(k-1)}&A_{(k)}     & \\
 &       &       &       &      &&U_{R}      &U_{R} \\\hline
  &       &      U_{L}&A_{(3)}&\cdots&A_{(k-1)}& A_{(k)}/U_{R} &U_{R}\\
 &P_{L}&      P_{L}&       &      &&           & \\\hline
 &       &      U_{L}&A_{(3)}&\cdots&A_{(k-1)}& A_{(k)}/U_{R} &U_{R}\\
 &P_{L}&      P'_{L}&       &      &&           & \\\hline
  &       &      U'_{L}&A_{(3)}&\cdots&A_{(k-1)}& A_{(k)}/U_{R} &U_{R}\\
 &P_{L}&      P_{L}&       &      &&           & \\\hline
 &A_{(1)}&A_{(2)}&A_{(3)}&\cdots&A_{(k-1)}& A_{(k)}/U_{R} &U_{R}
\end{array}
\end{equation*}
Here, the primed operator indicates that $\xu'=\yu$, $\yu'=\xu$, and is zero for all other operators. We choose $U_R=\xo$ if $A_{(k)}\in \{\yo,\zo\}$ and $U_R=\yo$ if $A_{(k)}=\xo$. These commutators imply the following equation:
\begin{align*}
0 & = q(\bm{A}) \coeff([A_{(k)},U_R]) J_\theta(U_R) \\
&+ J_\theta(P_L P_L) \coeff([U_L,P_L]|A_{(2)}) q(U_L A_{(3)}\cdots A_{(k-1)} A_{(k)}/U_R U_R) \\
&+ J_\theta(P_L P'_L) \coeff([U_L,P'_L]|A_{(2)}) q(U_L A_{(3)}\cdots A_{(k-1)} A_{(k)}/U_R U_R) \\
&+ J_\theta(P_L P_L) \coeff([U'_L,P_L]|A_{(2)}) q(U'_L A_{(3)}\cdots A_{(k-1)} A_{(k)}/U_R U_R).
\end{align*}
Here, $J_\theta(P_LP_L)$ or $J_\theta(P_LP'_L)$ means the coefficient of those operator string in the Hamiltonian $H_\theta$ in \ref{eq:Ham_modifiedform}, and $\coeff([U_L,P_L]|A_{(2)})$ represents the contribution of $A_{(2)}$ to the commutator $[U_L,P_L]$. Since $v\leq k-2$, the DP path of every operator string in this equation except $\bm{A}$ has at most $v-1$ vertices, and by the induction hypothesis, their coefficients must be zero. Thus, $q(\bm{A})=0$.

The case $A_{(k)}\in \{\xu,\yu\}$ follows similarly. Suppose $A_{(k)}=\xu$ and consider the following commutators:
\begin{equation*}
\begin{array}{ccccccccc}
 &A_{(1)}&A_{(2)}&A_{(3)}&\cdots&A_{(k-1)}&\xu     & \\
 &       &       &       &      &&\xo     &\xo \\\hline
  &A_{(1)}&A_{(2)}&A_{(3)}&\cdots&A_{(k-1)}&\yu     & \\
 &       &       &       &      &&\xo     &\xo \\\hline
  &       &      U_{L}&A_{(3)}&\cdots&A_{(k-1)}& X &\xo\\
 &P_{L}&      P_{L}&       &      &&           & \\\hline
 &       &      U_{L}&A_{(3)}&\cdots&A_{(k-1)}& X &\xo\\
 &P_{L}&      P'_{L}&       &      &&           & \\\hline
  &       &      U'_{L}&A_{(3)}&\cdots&A_{(k-1)}& X &\xo\\
 &P_{L}&      P_{L}&       &      &&           & \\\hline
 &A_{(1)}&A_{(2)}&A_{(3)}&\cdots&A_{(k-1)}& X &\xo
\end{array}
\end{equation*}
All operator strings in these commutators, except the first two, have zero coefficients due to the induction hypothesis, yielding $2q(\bm{A})  - q(A_{(1)}\cdots A_{(k-1)}\yu)=0$. Similarly, from the following commutators:
\begin{equation*}
\begin{array}{ccccccccc}
 &A_{(1)}&A_{(2)}&A_{(3)}&\cdots&A_{(k-1)}&\xu     & \\
 &       &       &       &      &&\yo     &\yo \\\hline
  &A_{(1)}&A_{(2)}&A_{(3)}&\cdots&A_{(k-1)}&\yu     & \\
 &       &       &       &      &&\yo     &\yo \\\hline
  &       &      U_{L}&A_{(3)}&\cdots&A_{(k-1)}& Y &\yo\\
 &P_{L}&      P_{L}&       &      &&           & \\\hline
 &       &      U_{L}&A_{(3)}&\cdots&A_{(k-1)}& Y &\yo\\
 &P_{L}&      P'_{L}&       &      &&           & \\\hline
  &       &      U'_{L}&A_{(3)}&\cdots&A_{(k-1)}& Y &\yo\\
 &P_{L}&      P_{L}&       &      &&           & \\\hline
 &A_{(1)}&A_{(2)}&A_{(3)}&\cdots&A_{(k-1)}& Y &\yo
\end{array}
\end{equation*}
we find $q(\bm{A})  - 2q(A_{(1)}\cdots A_{(k-1)}\yu)=0$. Combining these two equations gives $q(\bm{A}) = q(A_{(1)}\cdots A_{(k-1)}\yu)=0$. A similar argument applies when $A_{(k)}=\yu$.
\end{proof}

Now suppose that $\bm{A}$ has a DP path with $k-1$ vertices, but the operator in the last vertex of the DP path differs from $A_{(k)}$. This implies $\bm{A}$ satisfies
\begin{equation}\label{eq:onemiss}
\bm{A}\leq \ob{D_{(1)}\cdots D_{(k-1)}}/U_R.
\end{equation}
Here $D_{(n)}$ is an operator in the $n$-th vertex of the DP path, and $U_R$ is an operator such that $A_{(k)}\leq [D_{(k-1)},U_R]$. Notably, $U_R$ is always in the set $\{X,Y,Z\}$. For example, if $A_{(k)}=\xu$ and $D_{(k-1)}=\xo$, then we must choose $U_R=X$; similarly, if $A_{(k)}=\xo$ and $D_{(k-1)}=\yo$, then $U_R=Z$ is required. This implies that the approach we used in the proof of Lemma \ref{lemma:3} --- taking the commutator with the operator string $U_R U_R$ --- is not applicable here, as the coefficient of $U_R U_R$ in $H_{\arctan 2}$ is zero. However, by employing a strategy similar to that in Lemma \ref{lemma:4}, one can show that $q(\bm{A})$ is zero in this case.

To proceed, fix $U_R=X$; the cases $U_R=Y$ or $Z$ can be handled similarly. Now define $\bm{A}_{\text{base}}$ as $X/\ob{(\yo\xo)^{m-1}\yo}$ for even $k=2m$ and $X/\ob{(\yo\xo)^{m}}$ for odd $k=2m+1$.

\begin{lemma}
Suppose that $\bm{A}$ satisfies Eq.\ref{eq:onemiss} with $U_R=X$. Furthermore, suppose that the coefficient of $\bm{A}_{\text{base}}$ in $C$ is fixed. Then the coefficient $q(\bm{A})$ of $\bm{A}$ in $C$ is uniquely determined.
\end{lemma}
\begin{proof}
Suppose $A_{(k)}=\xo$, and consider the following commutators:
\begin{equation*}
\begin{array}{ccccccccc}
 &A_{(1)}&A_{(2)}&A_{(3)}&\cdots&A_{(k-1)}&\xo    & \\
 &       &       &       &      &&\yo       &\yo \\\hline
  &       &      U_{L}&A_{(3)}&\cdots&A_{(k-1)}&Z       &\yo  \\
 &P_{L}&      P_{L}&       &      &&      & \\\hline
 &       &      U_{L}&A_{(3)}&\cdots&A_{(k-1)}&Z       &\yo \\
 &P_{L}&      P'_{L}&       &      &&       & \\\hline
  &       &      U'_{L}&A_{(3)}&\cdots&A_{(k-1)}&Z       &\yo \\
 &P_{L}&      P_{L}&       &      &&       & \\\hline
 &A_{(1)}&A_{(2)}&A_{(3)}&\cdots&A_{(k-1)}&Z       &\yo 
\end{array}
\end{equation*}
If $U_L\not\in \{\xu, \yu\}$, then the last commutator does not contribute. Additionally, the second and third commutators both contain the operator string $U_LA_{(3)}\cdots A_{(k-1)}Z\yo$, creating a linear relationship between $q(U_LA_{(3)}\cdots A_{(k-1)}Z\yo)$ and $q(\bm{A})$. Notice that, due to Eq.\ref{eq:onemiss}, we can also express this as:
\begin{equation*}
U_LA_{(3)}\cdots A_{(k-1)}Z\yo \sim \ob{D_{(2)}\cdots D_{(k-1)}}/X/\yo\yo \sim \ob{D_{(2)}\cdots D_{(k-1)}\zo}/X,
\end{equation*}
showing that this new operator string still satisfies Eq.\ref{eq:onemiss}.

Next, consider the case $U_L = \xu$. Here, the third commutator does not contribute, and we focus on the following three commutators:
\begin{equation*}
\begin{array}{ccccccccc}
 &A_{(1)}&A_{(2)}&A_{(3)}&\cdots&A_{(k-1)}&\xo    & \\
 &       &       &       &      &&\yo       &\yo \\\hline
  &       &      \xu&A_{(3)}&\cdots&A_{(k-1)}&Z       &\yo  \\
 &P_{L}&      P_{L}&       &      &&      & \\\hline
  &       &      \yu&A_{(3)}&\cdots&A_{(k-1)}&Z       &\yo \\
 &P_{L}&      P_{L}&       &      &&       & \\\hline
 &A_{(1)}&A_{(2)}&A_{(3)}&\cdots&A_{(k-1)}&Z       &\yo 
\end{array}
\end{equation*}
This yields a linear equation containing the coefficients $q(\xu A_{(3)}\cdots A_{(k-1)}Z\yo)$, $q(\xu A_{(3)}\cdots A_{(k-1)}Z\yo)$, and $q(\bm{A})$. Next, consider the following commutators:
\begin{equation*}
\begin{array}{ccccccccc}
 &\xu&A_{(3)}&A_{(4)}&\cdots&A_{(k-1)}&Z&\yo    & \\
 &       &      & &       &      &&\xo       &\xo \\\hline
  &       &      U_2&A_{(4)}&\cdots&A_{(k-1)}&   Z    &Z &\xo \\
 &\xu&      \xu&       &&      &&      & \\\hline
  &       &      U_2&A_{(4)}&\cdots&A_{(k-1)}&Z       &Z&\xo \\
 &\xu&      \yu&       &&      &&       & \\\hline
 &\xu&A_{(3)}&A_{(4)}&\cdots&A_{(k-1)}&Z       &Z & \xo 
\end{array}
\end{equation*}
This provides a linear relationship between $q(U_2A_{(4)}\cdots A_{(k-1)}ZZ\xo)$ and $q(\xu A_{(3)}\cdots A_{(k-1)}Z\yo)$, as well as between $q(U_2A_{(4)}\cdots A_{(k-1)}ZZ\xo)$ and $q(\yu A_{(3)}\cdots A_{(k-1)}Z\yo)$. Here $U_2$ satisfies $A_{(3)}\leq [U_2, \xu]$(finding such $U_2\in \{\xo,\yo,\zo\}$ allowed by Eq.\ref{eq:onemiss}; indeed, $U_2=D_{(3)}$). By combining these relationships, we derive a linear relation between $q(U_2A_{(4)}\cdots A_{(k-1)}ZZ\xo)$ and $q(\bm{A})$. A similar argument holds for $U_L=\yu$. Again, this new operator string still satisfies Eq.\ref{eq:onemiss}.

The same process applies when $A_{(k)}=\yo$. For this case, consider the following commutators:
\begin{equation*}
\begin{array}{ccccccccc}
 &A_{(1)}&A_{(2)}&A_{(3)}&\cdots&A_{(k-1)}&\yo    & \\
 &       &       &       &      &&\xo       &\xo \\\hline
  &       &      U_{L}&A_{(3)}&\cdots&A_{(k-1)}&Z       &\xo  \\
 &P_{L}&      P_{L}&       &      &&      & \\\hline
 &       &      U_{L}&A_{(3)}&\cdots&A_{(k-1)}&Z       &\xo \\
 &P_{L}&      P'_{L}&       &      &&       & \\\hline
  &       &      U'_{L}&A_{(3)}&\cdots&A_{(k-1)}&Z       &\xo \\
 &P_{L}&      P_{L}&       &      &&       & \\\hline
 &A_{(1)}&A_{(2)}&A_{(3)}&\cdots&A_{(k-1)}&Z       &\xo 
\end{array}
\end{equation*}
It is notable that
\begin{equation*}
U_LA_{(3)}\cdots A_{(k-1)}Z\xo \sim \ob{D_{(2)}\cdots D_{(k-1)}}/X/\xo\xo \sim \ob{D_{(2)}\cdots D_{(k-1)}\xu}/X,
\end{equation*}
so the new operator string also satisfies Eq.\ref{eq:onemiss}.

As this process continues, the length of the substring consisting of $Z$ operators continues to increase. After sufficiently many repetitions, we eventually reach the opreator string $UZ\cdots Z\yo$ for even $k$ and $UZ\cdots Z\xo$ for odd $k$, which also satisfies Eq.\ref{eq:onemiss}. We can then show that these operator strings satisfy Eq.\ref{eq:onemiss} only when $U=\zo$, giving the operator string $\bm{A}_{\text{base}}$. This establishes a linear relationship between $q(\bm{A})$ and $q(\bm{A}_{\text{base}})$.

Next, suppose $A_{(k)}=\xu$ and consider the following commutators:
\begin{equation*}
\begin{array}{ccccccccc}
 &A_{(1)}&A_{(2)}&A_{(3)}&\cdots&A_{(k-1)}&\xu   & \\
 &       &       &       &      &&\xo       &\xo \\\hline
  &A_{(1)}&A_{(2)}&A_{(3)}&\cdots&A_{(k-1)}&\yu   & \\
 &       &       &       &      &&\xo       &\xo \\\hline
  &       &      U_{L}&A_{(3)}&\cdots&A_{(k-1)}&X       &\xo  \\
 &P_{L}&      P_{L}&       &      &&      & \\\hline
 &       &      U_{L}&A_{(3)}&\cdots&A_{(k-1)}&X       &\xo \\
 &P_{L}&      P'_{L}&       &      &&       & \\\hline
  &       &      U'_{L}&A_{(3)}&\cdots&A_{(k-1)}&X       &\xo \\
 &P_{L}&      P_{L}&       &      &&       & \\\hline
 &A_{(1)}&A_{(2)}&A_{(3)}&\cdots&A_{(k-1)}&X       &\xo
\end{array}
\end{equation*}
The operator strings in the last three commutators also satisfy Eq.\ref{eq:onemiss}:
\begin{equation*}
U^{(\prime)}_LA_{(3)}\cdots A_{(k-1)}X\xo \sim \ob{D^{(\prime)}_{(2)}\cdots D_{(k-1)}}/X/\xo\xo \sim \ob{D_{(2)}\cdots D_{(k-1)}\xu}/X.
\end{equation*}
Thus, by the previous argument, the coefficients of the operator strings in the last three commutators are uniquely determined when $q(\bm{A}_{\text{base}})$ is fixed. Consequently, we have
\begin{equation*}
2q(A_{(1)}\cdots A_{(k-1)}\xu)-q(A_{(1)}\cdots A_{(k-1)} \yu) = c_1 q(\bm{A}_{\text{base}})
\end{equation*}
for some constant $c_1$, and similarly,
\begin{equation*}
-q(A_{(1)}\cdots A_{(k-1)} \xu)+2q(A_{(1)}\cdots A_{(k-1)} \yu) = c_2 q(\bm{A}_{\text{base}})
\end{equation*}
for some constant $c_2$. Solving these equations, we find
\begin{align*}
q(A_{(1)}\cdots  A_{(k-1)}\xu)&= \frac{2c_1+c_2}{3}q(\bm{A}_{\text{base}}),\\
q(A_{(1)}\cdots A_{(k-1)} \yu)&= \frac{c_1+2c_2}{3}q(\bm{A}_{\text{base}}).
\end{align*}
Thus, $q(\bm{A})$ is uniquely determined once $q(\bm{A}_{\text{base}})$ is fixed. The same argument applies for operator strings $\bm{A}$ ending with $\yu$. This completes the proof.
\end{proof}

\begin{lemma}
The coefficient $q(\bm{A}_{\text{base}})$ in $C$ must be zero. Therefore, every coefficient $q(\bm{A})$ for $\bm{A}$ satisfying Eq.\ref{eq:onemiss} with $U_R=X$ is zero.
\end{lemma}
\begin{proof}
Let $k=2m$ and consider the following commutators:
\begin{equation*}
\begin{array}{ccccccccc}
 &\zo&Z&Z&\cdots&Z&\yo   & \\
 &       &       &       &      &&\xo       &\xo \\\hline
  &       &      \xu&Z&\cdots&Z&Z       &\xo  \\
 &\zo&      \zo&       &      &&      & \\\hline
  &       &      \yu&Z&\cdots&Z&Z       &\xo  \\
 &\zo&      \zo&       &      &&      & \\\hline
 &\zo&Z&Z&\cdots &Z       &Z&\xo
\end{array}
\end{equation*}
This shows that
\begin{equation*}
q(\bm{A}_{\text{base}})-q(\xu Z\cdots Z\xo) - q(\yu Z\cdots Z \xo)=0.
\end{equation*}
Now consider the following commutators:
\begin{equation*}
\begin{array}{ccccccccc}
 &\xu&Z&Z&\cdots&Z&\xo   & \\
 &       &       &       &      &&\yo       &\yo \\\hline
  &       &      \zo&Z&\cdots&Z&Z       &\yo  \\
 &\xu&      \xu&       &      &&      & \\\hline
  &       &      \zo&Z&\cdots&Z&Z       &\yo  \\
 &\xu&      \yu&       &      &&      & \\\hline
 &\xu&Z&Z&\cdots &Z       &Z&\yo
\end{array}
\end{equation*}
This shows that $-J_3 q(\xu Z\cdots Z\xo) + 3J_2 q(\bm{A}_{\text{base}}) = 0$. Similarly, $-J_3 q(\yu Z\cdots Z\xo) + 3J_2 q(\bm{A}_{\text{base}}) = 0$. Combining all these equations, we get:
\begin{equation*}
(1-\frac{6J_2}{J_3})q(\bm{A}_{\text{base}})=0.
\end{equation*}
Since $\frac{J_2}{J_3} = \frac{2}{3}$, this simplifies to $q(\bm{A}_{\text{base}})=0$.
\end{proof}

Note that Lemmas \ref{lemma:5} and Lemma \ref{lemma:6} hold independently of the values of $J_1, J_2,$ and $J_3$. Therefore, we have shown that if $J_2,J_3\neq 0$ and $J_1=0$, i.e., $\theta=\arctan 2$, then $\len ([C,H_0])\leq \len C$ implies that $C|_k= \alpha C_{\text{boost}}$.

\section{Step 2: Detailed Proof}\label{section:4}
In this section, we prove Step 2 given in the main text for general $k\geq 3$. We first address the case where $J_1\neq 0$, followed by the case where $J_1=0$. Unlike the previous section, the proofs for these two cases do not differ significantly.

\subsection{Case 1: $J_1\neq 0$}
In proving our statement, the commutator between a doubling product operator and the operator string in $H_\theta$ appears frequently. To simplify these expressions, we introduce the following notation:
\begin{equation}
\{\ob{A_1\cdots A_{m-1}\underset{B}{A_m}A_{m+1}\cdots A_{n}}\}_j \coloneq c\cdot [\{\ob{A_1\cdots A_n}\}_j, \{BB\}_{j+m-1}]
\end{equation}
Here $A$ and $B$ represent one of the operators $X, Y, Z, \xo, \yo, \zo, \xu,\yu,\zu$, and $c$ is a normalization coefficient. The following lemmas use this notation. Note that the subscript in each Lemma is an index and does not indicate the position of the operator.

\begin{lemma}\label{lemma:15}
Let the operators $A_t$ and $B_t$ be one of the operators $X, Y, Z$. With appropriate choices of the operators $U$ and $P$, the following relations hold:
\begin{align*}
\ob{A_1 A_2\cdots A_m}/\yu\ob{B_1 B_2 B_3\cdots B_{n-1} B_n} &\leq \ob{A_1\cdots A_m \underset{\tcr{P}}{{U}}B_1B_2B_3\cdots B_{n-1} B_n}\\
''&\leq \ob{A_1\cdots A_m \yu\underset{\tcr{P}}{{U}}B_2B_3\cdots B_{n-1}B_n}\\
''&\leq \ob{A_1\cdots A_m \yu\overline{B_1}\underset{\tcr{P}}{{U}}B_3\cdots B_{n-1}B_n}\\
''&\leq\ob{A_1\cdots A_m \yu\overline{B_1}\overline{B_2}\underset{\tcr{P}}{{U}}\cdots B_{n-1}B_n}\\
&\vdots\\
''&\leq \ob{A_1\cdots A_m \yu\overline{B_1}\overline{B_2}\overline{B_3}\cdots \overline{B_{n-1}}\underset{\tcr{P}}{U}}\\
''&\leq \ob{\tcr{A_1}}/\ob{A_2\cdots A_m}\yu|\ob{B_1 B_2 B_3\cdots B_{n-1} B_n}\\
''&\leq \ob{A_1A_2\cdots A_m}\yu|\ob{B_1 B_2 B_3\cdots B_{n-1}}/\ob{\tcr{B_n}}
\end{align*}
Similarly, the following relations hold:
\begin{align*}
\ob{A_1 A_2\cdots A_m}/\yu\ob{B_1}&\leq \ob{A_1\cdots A_m \underset{\tcr{P}}{{U}} B_1}\\
''&\leq \ob{A_1\cdots A_m \yu \underset{\tcr{P}}{{U}}}\\
''&\leq \ob{\tcr{A_1}}/ \ob{A_2\cdots A_m}S|\ob{B_1}
\end{align*}
Finally, the following relations hold:
\begin{align*}
\yu\ob{X B_2\cdots B_{n-1}B_n} &\leq \ob{\underset{\tcr{P}}{{U}}XB_2\cdots B_{n-1}B_n}\\
''&\leq\ob{\yu\underset{\tcr{P}}{{U}}B_2\cdots B_{n-1}B_n}\\
''&\leq\yu\ob{X B_2\cdots B_{n-1}}/\ob{\tcr{B_n}}
\end{align*}
Here, the operators highlighted in red represent an operator string from the Hamiltonian $H_\theta$. These are the only ways to express these operators through the commutator between a length $k$ doubling product operator and the operator string from $H_\theta$, or the commutator between a length $k-1$ doubling product operator and the operator string from $H_\theta$.
\end{lemma}
This Lemma can be directly verified by calculating the commutators.

Next, we calculate the contribution of the commutator to the leftward operator in the equations of Lemma \ref{lemma:15}. Notice that, because $U$ and $P$ are not yet determined in these equations, treating each case individually can be complex. However, we can significantly reduce the number of calculations. For example, consider the following setup:
\begin{equation}\label{eq:loc1}
\begin{tikzpicture}
\draw (0,-0.5) node {$X$};
\draw (0.5,-0.5) node {$Z$};
\draw (1,-0.5) node {$Z$};
\draw (1.5,-0.5) node {$Z$};
\draw (2.5,-0.5) node {$\yo$};
\draw (3.5,-0.5) node {$X$};
\draw (4.5,-0.5) node {$Y$};
\draw (5,-0.5) node {$Y$};
\draw (5.5,-0.5) node {$Y$};
\draw (6,-0.5) node {$Z$};

\draw (2.5,0) node {$\tcb{P}$};
\draw (3.5,0) node {$\tcb{P}$};

\draw (-0.5,-0.25) -- (6.5,-0.25);

\draw (0,0.5) node {$X$};
\draw (0.5,0.5) node {$Z$};
\draw (1,0.5) node {$Z$};
\draw (1.5,0.5) node {$Z$};
\draw (2.5,0.5) node {$[Y,\tcr{U}]$};
\draw (3.5,0.5) node {$[\tcr{U},X]$};
\draw (4.5,0.5) node {$Y$};
\draw (5,0.5) node {$Y$};
\draw (5.5,0.5) node {$Y$};
\draw (6,0.5) node {$Z$};

\draw[line width = 0.5mm, green!70!black] (2,-0.75) rectangle (4,0.75);
\end{tikzpicture}
\end{equation}
Here $U,P\in \mathcal{B}$ are single-site operators. This configuration represents the commutator between $\ob{(XY)^{(2)}U(XZ)^{(2)}}$ and $\ob{P}$, which contributes to $\ob{XYXY}/\yu \ob{XZXZ}=\bm{B}$. The contribution $G$ depends only on the local information near the $U$ operators, as highlighted in the green box:
\begin{equation}
G(\ob{(XY)^{(2)}U(XZ)^{(2)}};\bm{B}) = G(\ob{\cdots YUX\cdots};\cdots \ob{\cdots Y}/\yu \ob{X\cdots}).
\end{equation}
Information outside the green box, denoted by $\cdots$, is not needed to compute $G$. Thus, we can abbreviate this as $G(\ob{YUX};\ob{Y}/\yu \ob{X})$. Therefore, the contribution of the doubling product operators $\ob{(XY)^{(2)}U(XZ)^{(2)}}$ to $\ob{XYXY}/\yu \ob{XZXZ}=\bm{B}$ becomes:
\begin{equation}\label{eq:loc_2}
\sum_{U,P\in \mathcal{B}} G(\ob{YUX};\ob{Y}/\yu \ob{X}) \gamma_\theta(\ob{(XY)^{(2)}U(XZ)^{(2)}}).
\end{equation}
By the definition of $\gamma_\theta$, we have:
\begin{align*}\label{eq:loc_3}
&\gamma_\theta(\ob{(XY)^{(2)}U(XZ)^{(2)}}) \\
=&\sum_{U'}\gamma_\theta(\ob{(XY)^{(2)}}) \coeff([Y,U'])J(U')\coeff([U',X]) \gamma_\theta(\ob{(XZ)^{(2)}})\numberthis
\end{align*}
where the summation runs over all $U'$ that satisfy $\ob{(XY)^{(2)}U(XZ)^{(2)}} = \ob{(XY)^{(2)}U'(XZ)^{(2)}}$. Thus, Eq.\ref{eq:loc_2} becomes:
\begin{equation}\label{eq:loc_4}
\gamma_\theta(\ob{(XY)^{(2)}}) F(\ob{Y\underline{\phantom{U}}X}\rightarrow \ob{Y}/\yu \ob{X}) \gamma_\theta(\ob{(XZ)^{(2)}})
\end{equation}
where:
\begin{align*}\label{eq:loc_5}
&F(\ob{Y\underline{\phantom{U}}X}\rightarrow \ob{Y}/\yu \ob{X})\\
\coloneqq &\sum_{U,P\in \mathcal{B}} G(\ob{YUX};\ob{Y}/\yu \ob{X})\coeff([Y,U])J(U)\coeff([U,X]).\numberthis
\end{align*}
Thus, we can factorize the contribution of specific doubling product operators into two distinct parts; (i) the part determined directly the operator string we want to achieve (outside the green box in Eq.\ref{eq:loc1}, as well as the $\gamma_\theta$ terms in Eq.\ref{eq:loc_4}) and (ii) the part determined by the operator strings in $H_\theta$(inside the green box in Eq.\ref{eq:loc1}, and the $F$ terms in Eq.\ref{eq:loc_4}).

By separating the $F$ terms from the $\gamma_\theta$

In summary, by separating the $F$ terms and $\gamma_\theta$ terms in the commutators within Lemma \ref{lemma:15}, we can clearly calculate how the rightward commutators contribute to the leftward operator strings.

The following provides a list of relations along with their contributions. For simplicity, the contribution factor $\alpha$ is omitted.
\begin{align*}
\yu\ob{XZXZ\cdots XZ} &\leq \ob{\underset{\tcr{P}}{{U}}XZXZ\cdots XZ}&F(\ob{\underline{\phantom{U}}X}\rightarrow \yu \ob{X})(-1)^{k/2+1}J_1^{k-2}\\
''&\leq\ob{\yu\underset{\tcr{P}}{{U}}ZXZ\cdots XZ}& F(\ob{\yu \underline{\phantom{U}} Z} \rightarrow \yu\ob{XZ})(-1)^{k/2}J_1^{k-3}J_2\\
''&\leq\yu\ob{XZXZ\cdots X}/\ob{\tcr{Z}}& -J_1 q(\yu\ob{XZXZ\cdots X})\\
\end{align*}
\begin{align*}
\ob{Y}/\yu\ob{XZXZ\cdots ZX} &\leq \ob{Y\underset{\tcr{P}}{{U}}XZXZ\cdots ZX} & F(\ob{Y\underline{\phantom{U}}X}\rightarrow \ob{Y}/\yu \ob{X})(-1)^{k/2}J_1^{k-2}\\
''&\leq \ob{Y\yu\underset{\tcr{P}}{{U}}ZXZ\cdots ZX}& F(\ob{\yu \underline{\phantom{U}} Z} \rightarrow \yu\ob{XZ}) (-1)^{k/2+1} J_1^{k-3}J_2\\
''&\leq \ob{Y\yu\xo\underset{\tcr{P}}{{U}}XZ\cdots ZX}& F(\ob{\xo \underline{\phantom{U}} X} \rightarrow \ob{XZX})(-1)^{k/2} J_1^{k-4}J_2J_3\\
''&\leq \ob{Y\yu\xo\zo\underset{\tcr{P}}{{U}}Z\cdots ZX} & F(\ob{\zo \underline{\phantom{U}} Z} \rightarrow \ob{ZXZ})(-1)^{k/2}J_1^{k-5}J_2J_3^2\\
&\vdots\\
''&\leq \ob{Y\yu\xo\zo\xo\zo \cdots \underset{\tcr{P}}{{U}}X} & F(\ob{\xo \underline{\phantom{U}} X}\rightarrow \ob{XZX})(-1)^{k/2}J_1^2 J_2 J_3^{k-5}\\
''&\leq \ob{Y\yu\xo\zo\xo\zo \cdots \zo\underset{\tcr{P}}{{U}}} & F(\ob{\zo \underline{\phantom{U}}}\rightarrow \ob{ZX})(-1)^{k/2+1} J_1 J_2 J_3^{k-4}\\
''&\leq \ob{Y}/\yu\ob{XZXZ\cdots Z}/\ob{\tcr{X}} & q(\ob{Y}/\yu\ob{XZXZ\cdots Z})J_1\\
''&\leq \ob{\tcr{Y}}/\yu\ob{XZXZ \cdots ZX}& -2 q(\yu  \ob{XZXZ\cdots ZX}) J_1\\
''&\leq \ob{\tcr{Y}}/\xu\ob{XZXZ \cdots ZX}& q(\xu  \ob{XZXZ\cdots ZX}) J_1\\
\end{align*}
\begin{align*}
\ob{XY}/\yu\ob{XZXZ\cdots XZ} &\leq \ob{XY\underset{\tcr{P}}{{U}}XZXZ\cdots XZ} & F(\ob{Y\underline{\phantom{U}}X}\rightarrow \ob{Y}/\yu\ob{X})(-1)^{k/2} J_1^{k-2}\\
''&\leq \ob{XY\yu\underset{\tcr{P}}{{U}}ZXZ\cdots XZ} & F(\ob{\yu \underline{\phantom{U}}Z}\rightarrow \yu\ob{XZ}) (-1)^{k/2+1} J_1^{k-3}J_2\\
''&\leq \ob{XY\yu\xo\underset{\tcr{P}}{{U}}XZ\cdots XZ}& F(\ob{\xo \underline{\phantom{U}} X }\rightarrow \ob{XZX}) (-1)^{k/2} J_1^{k-4} J_2 J_3\\
''&\leq \ob{XY\yu\xo\zo\underset{\tcr{P}}{{U}}Z\cdots XZ}& F(\ob{\zo \underline{\phantom{U}} Z} \rightarrow \ob{ZXZ}) (-1)^{k/2} J_1^{k-5}J_2J_3^2 \\
&\vdots\\
''&\leq \ob{XY\yu\xo\zo\xo\zo \cdots \underset{\tcr{P}}{{U}}Z} & F(\ob{\zo \underline{\phantom{U}}Z}\rightarrow \ob{ZXZ}) (-1)^{k/2}J_1^3 J_2 J_3^{k-6}\\
''&\leq \ob{XY\yu\xo\zo\xo\zo \cdots \xo \underset{\tcr{P}}{{U}}} & F(\ob{\xo \underline{\phantom{U}} }\rightarrow \ob{XZ}) (-1)^{k/2}J_1^2 J_2 J_3^{k-5}\\
''&\leq \ob{XY}/\yu\ob{XZXZ\cdots X}/\ob{\tcr{Z}} & -q(\ob{XY}/\yu\ob{XZXZ\cdots X}) J_1\\
''&\leq \ob{\tcr{X}}/\ob{Y}/\yu\ob{XZXZ \cdots XZ} & -q(\ob{Y}/\yu\ob{XZXZ\cdots XZ}) J_1\\
\end{align*}
\begin{align*}
\ob{YXY}/\yu\ob{XZXZ\cdots ZX} &\leq \ob{YXY\underset{\tcr{P}}{{U}}XZXZ\cdots ZX} & F(\ob{Y\underline{\phantom{U}}X}\rightarrow \ob{Y}/\yu\ob{X}) (-1)^{k/2}J_1^{k-2}\\
''&\leq \ob{YXY\yu\underset{\tcr{P}}{{U}}ZXZ\cdots ZX}& F(\ob{\yu \underline{\phantom{U}}Z}\rightarrow \yu\ob{XZ}) (-1)^{k/2+1} J_1^{k-3}J_2\\
''&\leq \ob{YXY\yu\xo\underset{\tcr{P}}{{U}}XZ\cdots ZX}& F(\ob{\xo \underline{\phantom{U}} X }\rightarrow \ob{XZX}) (-1)^{k/2} J_1^{k-4} J_2 J_3\\
''&\leq \ob{YXY\yu\xo\zo\underset{\tcr{P}}{{U}}Z\cdots ZX}& F(\ob{\zo \underline{\phantom{U}} Z} \rightarrow \ob{ZXZ}) (-1)^{k/2} J_1^{k-5}J_2J_3^2 \\
&\vdots\\
''&\leq \ob{YXY\yu\xo\zo\xo\zo \cdots \zo \underset{\tcr{P}}{{U}}}& F(\ob{\zo \underline{\phantom{U}}}\rightarrow \ob{ZX}) (-1)^{k/2+1} J_1^3 J_2 J_3^{k-6}\\
''&\leq \ob{YXY}/\yu\ob{XZXZ\cdots Z}/\ob{\tcr{X}} & q(\ob{YXY}/\yu\ob{XZXZ\cdots Z}J_1)\\
''&\leq \ob{\tcr{Y}}/\ob{XY}/\yu\ob{XZXZ \cdots ZX}& q(\ob{XY}/\yu\ob{XZXZ \cdots ZX})J_1\\
\end{align*}
A similar argument applies to the other operator strings, until we obtain
\begin{align*}
\ob{XYX\cdots XY}/\yu\ob{XZ} &\leq \ob{XYX\cdots XY\underset{\tcr{P}}{{U}}XZ} & F(\ob{Y\underline{\phantom{U}}X}\rightarrow \ob{Y}/\yu\ob{X})(-1)^{k/2}J_1^{k-2}\\
''&\leq \ob{XYX\cdots XY\yu\underset{\tcr{P}}{{U}}Z} & F(\ob{\yu \underline{\phantom{U}}Z}\rightarrow \yu\ob{XZ}) (-1)^{k/2+1}J_1^{k-3}J_2\\
''&\leq \ob{XYX\cdots XY\yu\xo \underset{\tcr{P}}{{U}}}& F(\ob{\xo \underline{\phantom{U}} }\rightarrow \ob{XZ}) (-1)^{k/2}J_1^{k-4}J_2J_3\\
''&\leq \ob{XYX\cdots XY}/\yu\ob{X}/\ob{\tcr{Z}} & -q(\ob{XYX\cdots XY}/\yu\ob{X})J_1\\
''&\leq \ob{\tcr{X}}/\ob{YX\cdots XY}/\yu\ob{XZ} & -q(\ob{YX\cdots XY}/\yu\ob{XZ})J_1
\end{align*}
and
\begin{align*}
\ob{YXYX\cdots XY}/\yu\ob{X} &\leq \ob{YXYX\cdots XY\underset{\tcr{P}}{{U}}X} & F(\ob{Y\underline{\phantom{U}}X}\rightarrow \ob{Y}/\yu\ob{X}) (-1)^{k/2} J_1^{k-2}\\
''&\leq \ob{YXYX\cdots XY\yu\underset{\tcr{P}}{{U}}}& F(\ob{\yu \underline{\phantom{U}}} \rightarrow \yu\ob{X}) (-1)^{k/2} J_1^{k-3} J_2\\
''&\leq \ob{\tcr{Y}}/\ob{XYX\cdots XY}/\yu\ob{X} & q(\ob{XYX\cdots XY}/\yu\ob{X}) J_1
\end{align*}

Adding them all together, we obtain
\begin{align*}
0&=(k-3) F(\ob{Y\underline{\phantom{U}}X}\rightarrow \ob{Y}/\yu\ob{X}) (-1)^{k/2}J_1^{k-2}\\
&+ (k-4) F(\ob{\yu \underline{\phantom{U}}Z}\rightarrow \yu\ob{XZ}) (-1)^{k/2+1} J_1^{k-3}J_2\\
&+ (k-5) F(\ob{\xo \underline{\phantom{U}} X }\rightarrow \ob{XZX}) (-1)^{k/2}J_1^{k-4}J_2J_3\\
&+ (k-6) F(\ob{\zo \underline{\phantom{U}} Z} \rightarrow \ob{ZXZ}) (-1)^{k/2}J_1^{k-5} J_2 J_3^2\\
&+ \vdots\\
&+ F(\ob{\xo \underline{\phantom{U}} X }\rightarrow \ob{XZX}) (-1)^{k/2} J_1^2 J_2 J_3^{k-5}\\
&+ F(\ob{\zo \underline{\phantom{U}}}\rightarrow \ob{ZX}) (-1)^{k/2+1}J_2 (J_1J_3^{k-4} + J_1^3 J_3^{k-6}+\cdots + J_1^{k-5}J_3^2)\\
&+F(\ob{\xo \underline{\phantom{U}} }\rightarrow \ob{XZ}) (-1)^{k/2} J_2 (J_1^2 J_3^{k-5}+ J_1^4 J_3^{k-7}+\cdots+J_1^{k-4}J_3)\\
&+ F(\ob{\yu \underline{\phantom{U}}} \rightarrow \yu\ob{X}) (-1)^{k/2} J_1^{k-3}J_2\\
&-2 q(\yu \ob{XZXZ\cdots ZX}) J_1\\
&+ q(\xu \ob{XZXZ\cdots ZX})J_1=0
\end{align*}

Similarly,
\begin{align*}
\ob{Z}/\zu\ob{XZXZ\cdots ZX} &= \ob{Z\underset{\tcr{P}}{{U}}XZXZ\cdots ZX} & F(\ob{Z\underline{\phantom{U}}X}\rightarrow \ob{Z}/\zu \ob{X})(-1)^{k/2}J_1^{k-2}\\
''&\leq \ob{Z\zu\underset{\tcr{P}}{{U}}ZXZ\cdots ZX}& F(\ob{\zu \underline{\phantom{U}} Z} \rightarrow \zu\ob{XZ}) (-1)^{k/2+1} J_1^{k-3}J_2\\
''&\leq \ob{Z\zu\xo\underset{\tcr{P}}{{U}}XZ\cdots ZX}& F(\ob{\xo \underline{\phantom{U}} X} \rightarrow \ob{XZX})(-1)^{k/2} J_1^{k-4}J_2J_3\\
''&\leq \ob{Z\zu\xo\zo\underset{\tcr{P}}{{U}}Z\cdots ZX} & F(\ob{\zo \underline{\phantom{U}} Z }\rightarrow \ob{ZXZ})(-1)^{k/2}J_1^{k-5}J_2J_3^2\\
&\vdots\\
''&\leq \ob{Z\zu\xo\zo\xo\zo \cdots \underset{\tcr{P}}{{U}}X} & F(\ob{\xo \underline{\phantom{U}} X}\rightarrow \ob{XZX})(-1)^{k/2}J_1^2 J_2 J_3^{k-5}\\
''&\leq \ob{Z\zu\xo\zo\xo\zo \cdots \zo\underset{\tcr{P}}{{U}}} & F(\ob{\zo \underline{\phantom{U}} }\rightarrow \ob{ZX})(-1)^{k/2+1} J_1 J_2 J_3^{k-4}\\
''&\leq \ob{Z}/\zu\ob{XZXZ\cdots Z}/\ob{\tcr{X}} & q(\ob{Z}/\zu\ob{XZXZ\cdots Z})J_1\\
''&\leq \ob{\tcr{Z}}/\yu\ob{XZXZ \cdots ZX}& q(\yu  \ob{XZXZ\cdots ZX}) J_1\\
''&\leq \ob{\tcr{Z}}/\xu\ob{XZXZ \cdots ZX}& q(\xu  \ob{XZXZ\cdots ZX}) J_1\\
\end{align*}
\begin{align*}
\ob{YZ}/\zu\ob{XZXZ\cdots XZ} &= \ob{YZ\underset{\tcr{P}}{{U}}XZXZ\cdots XZ} & F(\ob{Z\underline{\phantom{U}}X}\rightarrow \ob{Z}/\zu \ob{X})(-1)^{k/2} J_1^{k-2}\\
''&\leq \ob{YZ\zu\underset{\tcr{P}}{{U}}ZXZ\cdots XZ} &  F(\ob{\zu \underline{\phantom{U}} Z} \rightarrow \zu\ob{XZ}) (-1)^{k/2+1} J_1^{k-3}J_2\\
''&\leq \ob{YZ\zu\xo\underset{\tcr{P}}{{U}}XZ\cdots XZ}& F(\ob{\xo \underline{\phantom{U}} X} \rightarrow \ob{XZX}) (-1)^{k/2} J_1^{k-4} J_2 J_3\\
''&\leq \ob{YZ\zu\xo\zo\underset{\tcr{P}}{{U}}Z\cdots XZ}& F(\ob{\zo \underline{\phantom{U}} Z }\rightarrow \ob{ZXZ}) (-1)^{k/2} J_1^{k-5}J_2J_3^2 \\
&\vdots\\
''&\leq \ob{YZ\zu\xo\zo\xo\zo \cdots \underset{\tcr{P}}{{U}}Z} &F(\ob{\zo \underline{\phantom{U}} Z}\rightarrow \ob{ZXZ}) (-1)^{k/2}J_1^3 J_2 J_3^{k-6}\\
''&\leq \ob{YZ\zu\xo\zo\xo\zo \cdots \xo \underset{\tcr{P}}{{U}}} & F(\ob{\xo \underline{\phantom{U}} }\rightarrow \ob{XZ}) (-1)^{k/2}J_1^2 J_2 J_3^{k-5}\\
''&\leq \ob{YZ}/\zu\ob{XZXZ\cdots X}/\ob{\tcr{Z}} & -q(\ob{YZ}/\zu\ob{XZXZ\cdots X}) J_1\\
''&\leq \ob{\tcr{Y}}/\ob{Z}/\zu\ob{XZXZ \cdots XZ} & -q(\ob{Z}/\zu\ob{XZXZ\cdots XZ}) J_1\\
\end{align*}
\begin{align*}
\ob{ZYZ}/\zu\ob{XZXZ\cdots ZX} &= \ob{ZYZ\underset{\tcr{P}}{{U}}XZXZ\cdots ZX} & F(\ob{Z\underline{\phantom{U}}X}\rightarrow \ob{Z}/\zu \ob{X}) (-1)^{k/2}J_1^{k-2}\\
''&\leq \ob{ZYZ\zu\underset{\tcr{P}}{{U}}ZXZ\cdots ZX}& F(\ob{\zu \underline{\phantom{U}} Z} \rightarrow \zu\ob{XZ}) (-1)^{k/2+1} J_1^{k-3}J_2\\
''&\leq \ob{ZYZ\zu\xo\underset{\tcr{P}}{{U}}XZ\cdots ZX}& F(\ob{\xo \underline{\phantom{U}} X} \rightarrow \ob{XZX}) (-1)^{k/2} J_1^{k-4} J_2 J_3\\
''&\leq \ob{ZYZ\zu\xo\zo\underset{\tcr{P}}{{U}}Z\cdots ZX}& F(\ob{\zo \underline{\phantom{U}} Z }\rightarrow \ob{ZXZ}) (-1)^{k/2} J_1^{k-5}J_2J_3^2 \\
&\vdots\\
''&\leq \ob{ZYZ\zu\xo\zo\xo\zo \cdots \zo \underset{\tcr{P}}{{U}}}& F(\ob{\zo \underline{\phantom{U}} }\rightarrow \ob{ZX}) (-1)^{k/2+1} J_1^3 J_2 J_3^{k-6}\\
''&\leq \ob{ZYZ}/\zu\ob{XZXZ\cdots Z}/\ob{\tcr{X}} & q(\ob{ZYZ}/\zu\ob{XZXZ\cdots Z}J_1)\\
''&\leq \ob{\tcr{Z}}/\ob{YZ}/\zu\ob{XZXZ \cdots ZX}& q(\ob{YZ}/\zu\ob{XZXZ \cdots ZX})J_1\\
\end{align*}
A similar argument applies to the other operator strings, until we obtain
\begin{align*}
\ob{YZY\cdots YZ}/\zu\ob{XZ} &= \ob{YZY\cdots YZ\underset{\tcr{P}}{{U}}XZ} &  F(\ob{Z\underline{\phantom{U}}X}\rightarrow \ob{Z}/\zu \ob{X})(-1)^{k/2}J_1^{k-2}\\
''&\leq \ob{YZY\cdots YZ\zu\underset{\tcr{P}}{{U}}Z} & F(\ob{\zu \underline{\phantom{U}} Z} \rightarrow \zu\ob{XZ}) (-1)^{k/2+1}J_1^{k-3}J_2\\
''&\leq \ob{YZY\cdots YZ\zu\xo \underset{\tcr{P}}{{U}}}& F(\ob{\xo \underline{\phantom{U}}} \rightarrow \ob{XZ}) (-1)^{k/2}J_1^{k-4}J_2J_3\\
''&\leq \ob{YZY\cdots YZ}/\zu\ob{X}/\ob{\tcr{Z}} & -q(\ob{YZY\cdots YZ}/\zu\ob{X})J_1\\
''&\leq \ob{\tcr{Y}}/\ob{ZY\cdots YZ}/\zu\ob{XZ} & -q(\ob{ZY\cdots YZ}/\zu\ob{XZ})J_1
\end{align*}
and
\begin{align*}
\ob{ZYZY\cdots YZ}/\zu\ob{X} &= \ob{ZYZY\cdots YZ\underset{\tcr{P}}{{U}}X} &  F(\ob{Z\underline{\phantom{U}}X}\rightarrow \ob{Z}/\zu \ob{X}) (-1)^{k/2} J_1^{k-2}\\
''&\leq \ob{ZYZY\cdots YZ\zu\underset{\tcr{P}}{{U}}}& F(\ob{\zu \underline{\phantom{U}}} \rightarrow \zu\ob{X}) (-1)^{k/2} J_1^{k-3} J_2\\
''&\leq \ob{\tcr{Z}}/\ob{YZY\cdots YZ}/\zu\ob{X} & q(\ob{YZY\cdots YZ}/\zu\ob{X}) J_1
\end{align*}

Adding them all together, we obtain
\begin{align*}
0&=(k-3) F(\ob{Z\underline{\phantom{U}}X}\rightarrow \ob{Z}/\zu \ob{X}) (-1)^{k/2}J_1^{k-2}\\
&+ (k-4) F(\ob{\zu \underline{\phantom{U}} Z} \rightarrow \zu\ob{XZ}) (-1)^{k/2+1} J_1^{k-3}J_2\\
&+ (k-5) F(\ob{\xo \underline{\phantom{U}} X} \rightarrow \ob{XZX}) (-1)^{k/2}J_1^{k-4}J_2J_3\\
&+ (k-6) F(\ob{\zo \underline{\phantom{U}} Z }\rightarrow \ob{ZXZ}) (-1)^{k/2}J_1^{k-5} J_2 J_3^2\\
&+ \vdots\\
&+ F(\ob{\xo \underline{\phantom{U}} X} \rightarrow \ob{XZX}) (-1)^{k/2} J_1^2 J_2 J_3^{k-5}\\
&+ F(\ob{\zo \underline{\phantom{U}} }\rightarrow \ob{ZX}) (-1)^{k/2+1}J_2 (J_1J_3^{k-4} + J_1^3 J_3^{k-6}+\cdots + J_1^{k-5}J_3^2)\\
&+ F(\ob{\xo \underline{\phantom{U}} }\rightarrow \ob{XZ}) (-1)^{k/2} J_2 (J_1^2 J_3^{k-5}+ J_1^4 J_3^{k-7}+\cdots+J_1^{k-4}J_3)\\
&+ F(\ob{\zu \underline{\phantom{U}}} \rightarrow \zu\ob{X}) (-1)^{k/2} J_1^{k-3}J_2\\
&+ q(\yu  \ob{XZXZ\cdots ZX}) J_1\\
&+ q(\xu \ob{XZXZ\cdots ZX})J_1
\end{align*}

In total, we get
\begin{align*}
0&=(k-3) (F(\ob{Y\underline{\phantom{U}}X}\rightarrow \ob{Y}/\yu\ob{X})-F(\ob{Z\underline{\phantom{U}}X}\rightarrow \ob{Z}/\zu \ob{X})) (-1)^{k/2}J_1^{k-2}\\
&+ (k-4) (F(\ob{\yu \underline{\phantom{U}}Z}\rightarrow \yu\ob{XZ})-F(\ob{\zu \underline{\phantom{U}} Z} \rightarrow \zu\ob{XZ})) (-1)^{k/2+1} J_1^{k-3}J_2\\
&+ (F(\ob{\yu \underline{\phantom{U}}} \rightarrow \yu\ob{X}) -F(\ob{\zu \underline{\phantom{U}}} \rightarrow \zu\ob{X}) (-1)^{k/2} J_1^{k-3}J_2\\
&-3 q(\yu \ob{XZXZ\cdots ZX}) J_1
\end{align*}
and therefore, with the aid of the first commutator relation,
\begin{align*}
0&=(k-3) (F(\ob{Y\underline{\phantom{U}}X}\rightarrow \ob{Y}/\yu\ob{X})-F(\ob{Z\underline{\phantom{U}}X}\rightarrow \ob{Z}/\zu \ob{X})) (-1)^{k/2}J_1^{k-2}\\
&+ (k-4) (F(\ob{\yu \underline{\phantom{U}}Z}\rightarrow \yu\ob{XZ})-F(\ob{\zu \underline{\phantom{U}} Z} \rightarrow \zu\ob{XZ})) (-1)^{k/2+1} J_1^{k-3}J_2\\
&+ (F(\ob{\yu \underline{\phantom{U}}} \rightarrow \yu\ob{X}) -F(\ob{\zu \underline{\phantom{U}}} \rightarrow \zu\ob{X}) (-1)^{k/2} J_1^{k-3}J_2\\
&-3 F(\ob{\underline{\phantom{U}}X}\rightarrow \yu \ob{X})(-1)^{k/2+1} J_1^{k-2}\\
&-3 F(\ob{\yu \underline{\phantom{U}} Z} \rightarrow \yu\ob{XZ})(-1)^{k/2}J_1^{k-3}J_2
\end{align*}

Since we have
\begin{gather*}
F(\ob{Y\underline{\phantom{U}}X}\rightarrow \ob{Y}/\yu\ob{X}) = -F(\ob{Z\underline{\phantom{U}}X}\rightarrow \ob{Z}/\zu \ob{X})) =\frac{(J_1+J_3)^2}{2} -\frac{J_2^2}{2}+2(J_1-J_3)J_3\\
F(\ob{\yu \underline{\phantom{U}}Z}\rightarrow \yu\ob{XZ}) = (J_1-J_3)(J_1+\frac{J_2}{2})\\
F(\ob{\zu \underline{\phantom{U}} Z} \rightarrow \zu\ob{XZ})) = (J_1-J_3) (-\frac{J_1}{2}+\frac{J_2}{2}+\frac{3}{2}J_3)\\
F(\ob{\yu \underline{\phantom{U}}} \rightarrow \yu\ob{X}) = -F(\ob{\zu \underline{\phantom{U}}} \rightarrow \zu\ob{X}) = -\frac{3}{2}(J_1-J_3)^2\\
F(\ob{\underline{\phantom{U}}X}\rightarrow \yu \ob{X}) = \frac{2}{3}J_1^2+4J_1J_3-2J_3^2-\frac{2}{3}J_2^2\\
F(\ob{\yu \underline{\phantom{U}} Z} \rightarrow \yu\ob{XZ}) = \frac{J_1^2}{2}+\frac{1}{2}J_2 J_3 - J_1J_3\\
\end{gather*}
we get
\begin{align*}
0&=(k-3) ((J_1+J_3)^2-J_2^2+4(J_1-J_3)J_3) (-1)^{k/2}J_1^{k-2}\\
&+ (k-4) \frac{3}{2}(J_1-J_3)^2 (-1)^{k/2+1} J_1^{k-3}J_2\\
&-3(J_1-J_3)^2 (-1)^{k/2} J_1^{k-3}J_2\\
&-3 (\frac{2}{3}J_1^2+4J_1J_3-2J_3^2-\frac{2}{3}J_2^2)(-1)^{k/2+1} J_1^{k-2}\\
&-3 (\frac{J_1^2}{2}+\frac{1}{2}J_2 J_3 - J_1J_3)(-1)^{k/2}J_1^{k-3}J_2
\end{align*}
Simplifying and substituting
\begin{align*}
J_1=\left(\cos \theta - \frac{\sin \theta}{2}\right),\quad J_2 = \sin\theta,\quad J_3=\frac{\sin\theta}{2}
\end{align*}
gives
\begin{equation}
\alpha (k-1) \cos\theta\cos 2\theta J_1^{k-3}=0
\end{equation}
Since $J_1\neq 0$, we may omit it. Define $f_k(\theta)\coloneqq (k-1) \cos\theta\cos 2\theta $. Since $k>2$, $f_k(\theta)\neq 0$ if and only if $\cos\theta\cos 2\theta\neq 0$, i.e., $\theta\not\in \{\pm \frac{\pi}{4}, \pm\frac{3\pi}{4},\pm\frac{\pi}{2}\}$. This implies $\alpha=0$ in this case, proving our claim.

\subsection{Case 2: $J_1=0$}

For the case $J_1=0$, it is not possible to directly apply our previous argument. However, due to the similarity between the operators $X,Y,Z$ and $\xo, \yo, \zo$, Lemma \ref{lemma:15} still holds when the operators $A_t, B_t$ are chosen from $xo, \yo, \zo$. Therefore, the proof for the $J_1\neq 0$ case can be readily adapted to the $J_1=0$ case by substituting $X, Y, Z$ in the doubling product form with $\xo, \yo, \zo$. In this case, we use $J_3$ in place $J_1$; hence, after detailed calculations, we find
\begin{equation}
\alpha(k-1) \cos\theta\cos 2\theta J_3^{k-3}=0.
\end{equation}
Since $J_1= 0$ we have $\theta = \arctan 2$, and this value shows that $\alpha=0$, proving our claim.

\section{Translationally noninvariant operators}
The argument treating about the translationally noninvariant operators, which we discuss in this section, follows the approach presented in \cite{park2024proof}. Suppose that, for the Hamiltonian $H_\theta$, there exists a translationally noninvariant local quantity $C$ of length $k>2$. Let $T^{(j)}$ denote the translation operator shifting by distance $j$. From $C$, we can define a translationally invariant local quantity $C_0$ as follows:
\begin{equation}
C_0:= \sum_{j=1}^{L} T^{(-j)} C T^{(j)}.
\end{equation}
Due to the translational invariance of $H_\theta$, $C_0$ is also conserved. Since we have eliminated the possibility of a translationally invariant local conserved quantity with length $k>2$, if the length of $C_0$ exceeds $2$, a contradiction arises, implying that such $C$ cannot exist. However, although $C$ is of length $k>2$, $C_0$ could still reduce to a conserved quantity with length $\leq 2$, such as $H_\theta, I,$ or $0$. We must therefore exclude this possibility.

Define the following quantities:
\begin{equation}
C_a = \sum_{j=1}^m e^{2\pi i a j/m }T^{(-j)}C T^{(j)},
\end{equation}
where $m$ is the smallest positive integer satisfying $T^{-m} C T^{m}=C$, which exists for finite $L$, and $a=1,2,\cdots, m-1$. Due to the translational invariance of $H_\theta$, $[C,H_\theta]=0$ implies $[C_a,H_\theta]=0$ for all $a$. Furthermore, the $C_a$ operators with length $k$ must exist since $mC=C_0+\sum_{a=1}^{m-1} C_a$, and $C$ is a length $k$ operator.

Our goal is to show that $[C,H]=0$ implies that every $C_a$ has a length less than $k$, contradicting above and proving that such $C$ does not exist. First, all arguments used in Section \ref{section:3} remain valid for $C_a$; the only difference is the factors $e^{2\pi i a j/m}$ multiplying the coefficients due to the phase difference between $q(\{\bm{A}\}_1)$ and $q(\{\bm{A}\}_{j+1})$ in $C_a$. These factors only modify the phase of the coefficient in the linear relation and do not affect the existence of the relation, so the arguments in Section \ref{section:3} hold.

Additionally, the arguments in Section \ref{section:4} are not required here. For $J_1\neq 0$ and $k=6m+1$, note that $q(\{\ob{(XY)^{3m}}\}_1)$ and $q(\{\ob{(XYZ)^{2m}}\}_{6m+1})$ are linearly related, as can be demonstrated using similar methods to those in Section \ref{section:3}. We also have
\begin{equation*}
q(\{\ob{(XY)^{3m}}\}_1) = -q(\{\ob{(YX)^{3m}}\}_2) = q(\{\ob{(XY)^{3m}}\}_3)
\end{equation*}
and since $q(\{\ob{(XY)^{3m}}\}_3)=e^{4\pi i a/m}q(\{\ob{(XY)^{3m}}\}_1)$, it follows that $a/m=1/2$. However,
\begin{equation*}
q(\{\ob{(XYZ)^{2m}}\}_1) = q(\{\ob{(YZX)^{2m}}\}_{6m+2})) = q(\{\ob{(ZXY)^{2m}}\}_{6m+3})) = q(\{\ob{(XYZ)^{2m}}\}_{6m+4}))
\end{equation*}
and $q(\{\ob{(XYZ)^{2m}}\}_{6m+4}))=e^{6\pi i a/m}q(\{\ob{(XYZ)^{2m}}\}_{6m+1}))$ imply $a/m=1/3$ or $a/m=2/3$, leading to a contradiction. Thus, $q(\{\ob{(XY)^{3m}}\}_1)=0$, and all other coefficients must be zero. A similar argument applies for any $k$ by adjusting operators in the doubling product form (e.g., using $\ob{(XY)^{3m}X}$ and $\ob{(XYZ)^{3m}X}$ for $k=6m+2$.), and for the $J_1=0$ case by substituting $\xo, \yo, \zo$ for $X, Y, Z$.

\section{Multiplication and Commutator tables}\label{section:7}
In this section, we present the multiplication and commutator tables for the single-site operators. The multiplication table for these operators is shown in Table \ref{table:mult}. Each result is expressed in its simplest form. Using this multiplication results, we derive the commutator table for the single-site operators, as shown in Table \ref{table:comm}. In this table, the imaginary unit $i$ has been omitted, and empty cells denote trivial commutators.

\begin{table}[htb]
\begin{adjustwidth}{-110pt}{-110pt}
\centering
\begin{tabular}{|c||c|c|c||c|c|c||c|c|c|}
\hline
$  \times $    & $X$         & $Y$         & $Z$         & $\underline{X}$& $\underline{Y}$& $\underline{Z}$& $\overline{X}$      & $\overline{Y}$      & $\overline{Z}$      \\ \hhline {|=||=|=|=||=|=|=||=|=|=|}
$X$    & $X^2$         & $\frac{\overline{Z}+iZ}{2}$         & $\frac{\overline{Y}-iY}{2}$        & $i\overline{X}$& $-\frac{X+i\overline{X}}{2}$&$\frac{X-i\overline{X}}{2}$& $-i\underline{X}$      & $\frac{Z-i\overline{Z}}{2}$& $\frac{Y+i\overline{Y}}{2}$     \\ \hline
$Y$    & $\frac{\overline{Z}-iZ}{2}$        & $Y^2$         & $\frac{\overline{X}+iX}{2}$         & $\frac{Y-i\overline{Y}}{2}$& $i\overline{Y}$& $-\frac{Y+i\overline{Y}}{2}$& $\frac{Z+i\overline{Z}}{2}$     & $-i\underline{Y}$      & $\frac{X-i\overline{X}}{2}$\\ \hline
$Z$    & $\frac{\overline{Y}+iY}{2}$         & $\frac{\overline{X}-iX}{2}$        & $Z^2$         &$ -\frac{Z+i\overline{Z}}{2}$& $\frac{Z-i\overline{Z}}{2}$& $i\overline{Z}$ & $\frac{Y-i\overline{Y}}{2}$& $\frac{X+i\overline{X}}{2}$     & $-i\underline{Z}$      \\ \hhline {|=||=|=|=||=|=|=||=|=|=|}
$\underline{X}$  & $-i\overline{X}$         & $\frac{Y+i\overline{Y}}{2}$      & $-\frac{Z-i\overline{Z}}{2}$      & $X^2$& $Z^2-I$& $Y^2-I$ & $iX$         & $\frac{\overline{Y}-iY}{2}$         & $-\frac{\overline{Z}+iZ}{2}$        \\ \hline
$\underline{Y}$  & $-\frac{X-i\overline{X}}{2}$     & $-i\overline{Y}$         & $\frac{Z+i\overline{Z}}{2}$      & $Z^2-I$&$Y^2$&  $X^2-I$  & $-\frac{\overline{X}+iX}{2}$        & $iY$         & $\frac{\overline{Z}-iZ}{2}$         \\ \hline
$\underline{Z}$  & $\frac{X+i\overline{X}}{2}$      & $-\frac{Y-i\overline{Y}}{2}$     & $-i\overline{Z}$         &$Y^2-I$ &$X^2-I$ & $Z^2$ & $\frac{\overline{X}-iX}{2}$         & $-\frac{\overline{Y}+iY}{2}$        & $iZ$         \\ \hhline {|=||=|=|=||=|=|=||=|=|=|}
$\overline{X}$ & $i\underline{X}$     & $\frac{Z-i\overline{Z}}{2}$      & $\frac{Y+i\overline{Y}}{2}$ & $-iX$& $-\frac{\overline{X}-iX}{2}$& $\frac{\overline{X}+iX}{2}$ & $X^2$         & $-\frac{\overline{Z}+iZ}{2}$        & $-\frac{\overline{Y}-iY}{2}$         \\ \hline
$\overline{Y}$ & $\frac{Z+i\overline{Z}}{2}$ & $i\underline{Y}$     & $\frac{X-i\overline{X}}{2}$      & $\frac{\overline{Y}+iY}{2}$&$-iY$ & $-\frac{\overline{Y}-iY}{2}$&  $-\frac{\overline{Z}-iZ}{2}$         & $Y^2$         & $-\frac{\overline{X}+iX}{2}$        \\ \hline
$\overline{Z}$ & $\frac{Y-i\overline{Y}}{2}$      & $\frac{X+i\overline{X}}{2}$ & $i\underline{Z}$     & $-\frac{\overline{Z}-iZ}{2}$& $\frac{\overline{Z}+iZ}{2}$& $-iZ$& $-\frac{\overline{Y}+iY}{2}$        & $-\frac{\overline{X}-iX}{2}$        & $Z^2$         \\ \hhline {|=||=|=|=||=|=|=||=|=|=|}
\end{tabular}
\end{adjustwidth}
\caption{Multiplication table for the single-site operators.}
\label{table:mult}
\end{table}

\begin{table}[htb]
\begin{adjustwidth}{-40pt}{-40pt}
\centering
\begin{tabular}{|c||c|c|c||c|c|c||c|c|c|}
\hline
$  [,]/i $    & $X$         & $Y$         & $Z$         & $\underline{X}$& $\underline{Y}$& $\underline{Z}$& $\overline{X}$      & $\overline{Y}$      & $\overline{Z}$      \\ \hhline {|=||=|=|=||=|=|=||=|=|=|}
$X$    & $ $         & $Z$         & $-Y$        &  $2\overline{X}$& $-\overline{X}$&$-\overline{X}$& $-2\underline{X}$      & $-\overline{Z}$& $\overline{Y}$     \\ \hline
$Y$    & $-Z$        & $ $         & $X$         & $-\overline{Y}$& $2\overline{Y}$& $-\overline{Y}$& $\overline{Z}$     & $-2\underline{Y}$      & $-\overline{X}$\\ \hline
$Z$    & $Y$         & $-X$        & $ $        &$ -\overline{Z}$& $-\overline{Z}$& $2\overline{Z}$ & $-\overline{Y}$& $\overline{X}$     & $-2\underline{Z}$      \\ \hhline {|=||=|=|=||=|=|=||=|=|=|}
$\underline{X}$  & $-2\overline{X}$         & $\overline{Y}$      & $\overline{Z}$     & $ $& $ $& $ $ & $2X$         & $-Y$         & $-Z$        \\ \hline
$\underline{Y}$  & $\overline{X}$     & $-2\overline{Y}$         & $\overline{Z}$      & $ $&$ $&  $ $  & $-X$        & $2Y$         & $-Z$         \\ \hline
$\underline{Z}$  & $\overline{X}$      & $\overline{Y}$     & $-2\overline{Z}$         &$ $ &$ $ & $ $ & $-X$         & $-Y$        & $2Z$         \\ \hhline {|=||=|=|=||=|=|=||=|=|=|}
$\overline{X}$ & $2\underline{X}$     & $-\overline{Z}$      & $\overline{Y}$ & $-2X$& $X$& $X$ & $ $         & $-Z$        & $Y$         \\ \hline
$\overline{Y}$ & $\overline{Z}$ & $2\underline{Y}$     & $-\overline{X}$      & $Y$&$-2Y$ & $Y$&  $Z$         & $ $         & $-X$        \\ \hline
$\overline{Z}$ & $-\overline{Y}$      & $\overline{X}$ & $2\underline{Z}$     & $Z$& $Z$& $-2Z$& $-Y$        & $X$        & $ $         \\ \hhline {|=||=|=|=||=|=|=||=|=|=|}
\end{tabular}
\end{adjustwidth}
\caption{Commutator table for the single-site operators, with empty cells indicating a commutator value of zero.}
\label{table:comm}
\end{table}

\bibliographystyle{plain}
\bibliography{biblio}

